\documentclass[10pt, letterpaper, conference]{IEEEtran}
\usepackage{enumitem}
\usepackage{hyperref}
\hypersetup{                    
  colorlinks,
  linkcolor={black},
  citecolor={red!70!black},
  urlcolor={blue!70!black}
}

\ifCLASSOPTIONcompsoc
\usepackage[nocompress]{cite}
\else
\usepackage{cite}
\fi

\usepackage{amsmath}
\usepackage{graphicx}
\usepackage{amsmath}
\usepackage{amsthm}
\usepackage{amssymb}
\usepackage{url}
\usepackage{comment}
\usepackage{listings}
\usepackage{xcolor}
\usepackage[framemethod=tikz]{mdframed}
\usepackage{multirow}

\newtheorem{definition}{Definition}

\newcommand{\pubkey}[1]{\mathit{pk}(#1)}
\newcommand{\aenc}[2]{\mathit{aenc}_{#2}(#1)}
\newcommand{\adec}[2]{\mathit{adec}_{#2}(#1)}

\newcommand{\sign}[2]{\mathit{sign}_{#2}(#1) }

\newcommand{\skBank}{\ensuremath{\mathit{priv\Bank}}}
\newcommand{\skCard}{\ensuremath{\mathit{priv\Card}}}
\newcommand{\skCA}{\ensuremath{\mathit{privCA}}}
\newcommand{\pkBank}{\ensuremath{\mathit{pub\Bank}}}
\newcommand{\pkCard}{\ensuremath{\mathit{pub\Card}}}
\newcommand{\pkCA}{\ensuremath{\mathit{pubCA}}}

\newcommand{\cert}[2]{\mathit{cert}_{#2}(#1) }

\newcommand{\fact}[1]{\ensuremath{\mathsf{#1}}}
\newcommand{\trace}{\ensuremath{\alpha}}
\newcommand{\transaction}{\ensuremath{\mathit{t}}}
\newcommand{\literalstr}[1]{\ensuremath{\mathtt{'#1'}}}
\newcommand{\tracesOf}[1]{\ensuremath{traces(#1)}}

\newcommand{\makerule}[3]
{
	\left[ \begin{array}{@{}c@{}}#1\end{array}\right]\hspace{-1ex}-\hspace{-1ex} 
	\left[ 
	\begin{array}{@{}c@{}}#2\end{array}\right]\hspace{-1ex}\xrightarrow{}\hspace{-1ex}
	\left[ \begin{array}{@{}c@{}}#3\end{array}\right] 
}

\newcommand{\signature}{\Sigma}
\newcommand{\variables}{\mathcal{V}}
\newcommand{\term}{\mathcal{T}_{\signature}(\variables)}

\newcommand{\groundterm}{\mathcal{T}_{\signature}}
\newcommand{\sort}{\mathcal{S}}

\newcommand{\facts}{\ensuremath{\mathcal{F}}}
\newcommand{\rules}{\ensuremath{\mathcal{R}}}
\newcommand{\specialrules}{\mathcal{A}}
\newcommand{\powerset}[1]{\ensuremath{\mathcal{P}(#1)}}
\newcommand{\powermultiset}[1]{\ensuremath{\mathcal{M}(#1)}}

\newcommand{\Card}{\ensuremath{C}}
\newcommand{\Bank}{\ensuremath{B}}

\newcommand{\datafont}[1]{\text{#1}}
\newcommand{\cmdfont}[1]{\ensuremath{\mathsf{#1}}}
\newcommand{\hexfont}[1]{\ensuremath{\mathtt{#1}}}

\newcommand{\PDOL}{\datafont{PDOL}}

\newcommand{\UN}{\datafont{UN}}

\newcommand{\AIP}{\datafont{AIP}}

\newcommand{\ATC}{\datafont{ATC}}

\newcommand{\CID}{\datafont{CID}}

\newcommand{\IAD}{\datafont{IAD}}

\newcommand{\AC}{\datafont{AC}}

\newcommand{\nc}{\datafont{NC}}

\newcommand{\CTQ}{\datafont{CTQ}}

\newcommand{\Km}{\ensuremath{mk}}
\newcommand{\Ks}{\ensuremath{s}}

\newcommand{\MAC}[2]{\mathit{MAC}_{#2}(#1) }
\newcommand{\MACprime}[2]{\mathit{MAC}_{#2}'(#1) }

\definecolor{codegreen}{rgb}{0,0.6,0}
\definecolor{codegray}{rgb}{0.5,0.5,0.5}
\definecolor{codepurple}{rgb}{0.58,0,0.82}
\definecolor{backcolour}{rgb}{0.98,0.98,0.95}
\definecolor{codebrown}{rgb}{0.5,0.0,0.0}

\lstdefinelanguage{tamarin}{
  morekeywords={lemma, All, Ex, not, exists, trace, rule, In, Out, let},
  sensitive=false,
  morecomment=[l]{//},
  morestring=[b]',
  literate={@}{{{\color{codebrown}@{}}}}1%
           {=}{{{\color{codebrown}={}}}}1%
           {<}{{{\color{codebrown}<{}}}}1%
           {>}{{{\color{codebrown}>{}}}}1%
           {[}{{{\color{codebrown}[{}}}}1%
           {]}{{{\color{codebrown}]{}}}}1%
           {|}{{{\color{codebrown}|{}}}}1%
           {\#}{{{\color{codebrown}\#{}}}}1%
           {:}{{{\color{codebrown}:{}}}}1%
           {\&}{{{\color{codebrown}\&{}}}}1%
           {-}{{{\color{codebrown}-{}}}}1
}

\lstdefinelanguage{tlog}{
  morecomment=[l]{//}
}

\newcommand{\falsified}{{\color{red}$\times$}}
\newcommand{\verified}{{\color{codegreen}$\checkmark$}}

\newcommand{\etal}{\textit{et al.}}
\newcommand*\circled[1]{\tikz[baseline=(char.base)]{
            \node[shape=circle,draw,inner sep=2pt] (char) {#1};}}

\usepackage[disable]{todonotes}

\usepackage[acronym, nomain]{glossaries}

\makeglossaries

\newacronym{APDU}{APDU}{Application Protocol Data Unit}
\newacronym{AID}{AID}{Application Identifier}

\newacronym{PDOL}{PDOL}{Processing Data Object List}
\newacronym{CVM}{CVM}{Cardholder Verification Method}
\newacronym{CVMR}{CVMR}{Cardholder Verification Method Results}
\newacronym{TVR}{TVR}{Terminal Verification Results}
\newacronym{AIP}{AIP}{Application Interchange Profile}
\newacronym{UN}{UN}{Unpredictable Number}
\newacronym{AFL}{AFL}{Application File Locator}
\newacronym{PAN}{PAN}{Primary Account Number}
\newacronym{CDOL}{CDOL}{Card Risk Management Data Object List}
\newacronym{ODA}{ODA}{Offline Data Authentication}

\newacronym{SDA}{SDA}{Static Data Authentication}
\newacronym{SSAD}{SSAD}{Signed Static Authentication Data}
\newacronym{DDA}{DDA}{Dynamic Data Authentication}
\newacronym{DDOL}{DDOL}{Dynamic Data Object List}
\newacronym{SDAD}{SDAD}{Signed Dynamic Authentication Data}
\newacronym{CDA}{CDA}{Combined Dynamic Data Authentication}
\newacronym{CDCVM}{CDCVM}{Consumer Device CVM}

\newacronym{AC}{AC}{Application Cryptogram}
\newacronym{TC}{TC}{Transaction Cryptogram}
\newacronym{ARQC}{ARQC}{Authorization Request Cryptogram}
\newacronym{AAC}{AAC}{Application Authentication Cryptogram}
\newacronym{ATC}{ATC}{Application Transaction Counter}
\newacronym{CID}{CID}{Cryptogram Information Data}
\newacronym{ARC}{ARC}{Authorization Response Code}
\newacronym{ARPC}{ARPC}{Authorization Response Cryptogram}
\newacronym{IAD}{IAD}{Issuer Application Data}
\newacronym{CTQ}{CTQ}{Card Transaction Qualifiers}
\newacronym{TTQ}{TTQ}{Terminal Transaction Qualifiers}
\newacronym{NFC}{NFC}{Near Field Communication}
\newacronym{HCE}{HCE}{Host-based Card Emulation}
\newacronym{POS}{POS}{Point-Of-Sale}
\newacronym{TA}{TA}{Transaction Authorization}
\newacronym{PSE}{PSE}{Payment System Environment}

\newacronym{IAC}{IAC}{Issuer Action Code}
\newacronym{TAC}{TAC}{Terminal Action Code}

\newacronym{PK}{PK}{Public Key}
\newacronym{MAC}{MAC}{Message Authentication Code}
\newacronym{CA}{CA}{Certificate Authority}

\begin{document}

\title{The EMV Standard: Break, Fix, Verify}

\author{
\IEEEauthorblockN{David Basin, Ralf Sasse, and Jorge Toro-Pozo}
\IEEEauthorblockA{\itshape Department of Computer Science\\
ETH Zurich, Switzerland}
}

\maketitle

\begin{abstract}
EMV is the international protocol standard for smartcard payment and is used in 
over 9 billion cards worldwide. Despite the standard's advertised security, 
various issues have been previously uncovered, deriving from logical flaws that 
are hard to spot in EMV's lengthy and complex specification, running over 2,000 
pages.

We formalize a comprehensive symbolic model of EMV in Tamarin, a 
state-of-the-art protocol verifier. Our model is the first that supports a 
fine-grained analysis of all relevant security guarantees that EMV is intended 
to offer. We use our model to automatically identify flaws that lead to two 
critical attacks: one that defrauds the cardholder and a second that defrauds 
the merchant. First, criminals can use a victim's Visa contactless card to make 
payments for amounts that require cardholder verification, without knowledge of 
the card's PIN. We built a proof-of-concept Android application and 
successfully demonstrated this attack on real-world payment terminals. Second, 
criminals can trick the terminal into accepting an unauthentic offline 
transaction, which the issuing bank should 
later decline, after the criminal has walked away with the goods. This attack 
is possible for implementations following the standard, although we did not 
test it on actual terminals for ethical reasons. Finally, we propose and verify 
improvements to the standard that prevent these attacks, as well as any other 
attacks that violate the considered security properties. The proposed 
improvements can be easily implemented in the terminals and do not affect the 
cards in circulation.
\end{abstract}

\section{Introduction}\label{sec:intro}

EMV, named after its founders Europay, Mastercard, and Visa, is the worldwide 
standard for smartcard payment, developed in the mid 1990s. As 
of December 2019, more than 80\% of all card-present transactions globally use 
EMV, reaching up to 98\% in many European countries. Banks have a strong 
incentive to adopt EMV due to the \emph{liability shift}, which relieves banks 
using the standard from any liability from payment disputes. If the disputed 
transaction was authorized by a PIN then the consumer (EMV terminology for the 
payment-card customer) is held liable. If a paper signature was used instead, 
then the merchant is charged.

\subsection*{EMV: 20 Years of Vulnerabilities}

Besides the liability shift, EMV's global acceptance is also attributed to 
its advertised security. However, EMV's security has been challenged numerous 
times. Man-in-the-middle (MITM) attacks~\cite{MurdochDAB10}, card 
cloning~\cite{Heydt-BenjaminBFJO07,RolandL13}, downgrade 
attacks~\cite{RolandL13}, relay 
attacks~\cite{DrimerM07,FrancisHMM11,SportielloC13,ChothiaGRBT15,BocekKTS16}, 
and card 
skimming~\cite{BondCMSA14, EmmsAFHM14} are all examples of successful 
exploits of the standard's shortcomings. The MITM attack reported 
by Murdoch \etal{}~\cite{MurdochDAB10} is believed to have been used by 
criminals in 2010--11 in France and Belgium to carry out fraudulent 
transactions for ca. 600,000 Euros~\cite{FerradiGNT16}. The underlying flaw of 
Murdoch \etal{}'s attack is that the card's response to the terminal's offline 
PIN verification request is not authenticated.

Some of the security issues identified result from flawed implementations of 
the standard. Others stem from logical flaws whose repairs would require 
changes to the entire EMV infrastructure. Identifying such flaws is far from 
trivial due to the complexity of EMV's execution flow, which is highly flexible 
in terms of card authentication modes, cardholder verification methods, and 
online/offline authorizations.  This raises the question of how we can 
systematically explore all possible executions and improve the standard to 
avoid another twenty years of attacks.


\subsection*{Approach Taken: Break, Fix, Verify}

In this paper we focus on weakness of and improvements to the EMV protocol 
design. We present a formal, comprehensive model for the symbolic 
analysis of EMV's security. Our model is written in 
Tamarin~\cite{tamarin13,SchmidtMCB12}, a state-of-the-art verification 
tool that has been used to study numerous real-world protocols, including TLS 
1.3~\cite{CremersHHSM17} and 5G authentication~\cite{BasinDHRSS18}. Tamarin 
supports protocol verification in the presence of powerful adversaries and 
unboundedly many concurrent protocol sessions. 

Our model supports the analysis of all properties that must hold in any  EMV 
transaction. An informal description of the three most relevant properties is 
as follows:
\begin{enumerate}	
\item \emph{Bank accepts terminal-accepted transactions}: No transaction
accepted by the terminal can be declined by the bank.
\item \emph{Authentication to the terminal}: All transactions accepted by the 
terminal are authenticated by the card and, if authorized online, the bank.
\item \emph{Authentication to the bank}: All transactions accepted by the bank 
are authenticated by the card and the terminal.
\end{enumerate}

Our model faithfully considers the three roles present in an EMV session: the 
bank, the terminal, and the card. Previous symbolic models merge the terminal 
and the bank into a single agent~\cite{RuiterP11,MauwSTT18,DebantDW18}. This 
merging incorrectly entails that the terminal can verify all 
card-produced cryptographic proofs that the bank can. This is incorrect as 
the card and the bank share a symmetric key that is only known to them.

Using our model, we identify a critical violation of authentication properties 
by the Visa contactless protocol: the cardholder verification method used in 
a transaction, if any, is neither authenticated nor cryptographically protected 
against modification. We developed a proof-of-concept Android 
application that exploits this to \textbf{bypass PIN verification} by 
mounting a man-in-the-middle attack that instructs the 
terminal that PIN verification is not required because the cardholder 
verification was performed on the consumer's device (e.g., a mobile phone). 
This enables criminals to use any stolen Visa card to pay for 
expensive goods without the card's PIN. In other words, \emph{the PIN is 
useless in Visa contactless transactions!}

We have successfully tested our PIN bypass attack on real-world 
terminals for a number of transactions with amounts greater than the limit 
above 
which cardholder verification is required (which we will refer to as 
\emph{high-value} transactions), using Visa-branded 
cards such as Visa Credit, Visa Electron, and V Pay cards. For example, we 
performed a transaction of ca. \$190 in an attended terminal in 
an actual store. As it is now common for consumers to pay with their 
smartphones, the 
cashier cannot distinguish the attacker's actions from those of any legitimate 
cardholder. We carried out all our tests using our own cards, but we stress 
that the attack works for any Visa card 
that the attacker possesses, in particular with stolen cards.

Our symbolic analysis using Tamarin also reveals that, in an offline 
contactless transaction with a Visa or an old Mastercard card, the card does 
not authenticate to the terminal the \acrfull{AC}, which is a card-produced
cryptographic proof of the transaction that the terminal \emph{cannot} verify 
(only the card issuer can). This enables criminals to \textbf{trick the 
terminal into accepting an unauthentic offline transaction}. When 
the acquirer later submits the transaction data as part of the clearing record, 
the issuing bank will detect the wrong cryptogram, but the criminal is already 
long gone with the goods. We did not test this attack on actual terminals for 
ethical reasons as this would defraud the merchant.

\subsection*{Contributions}

First, we present a comprehensive symbolic model of the EMV standard 
that accounts for the three \acrlong{ODA} methods (\acrshort{SDA}, 
\acrshort{DDA}, and \acrshort{CDA}), the five \acrlong{CVM}s (no PIN, plaintext 
PIN, offline enciphered PIN, online PIN, and \acrshort{CDCVM}), the two types 
of 
\acrlong{TA}s (offline and online), and the two (major) types of contactless 
transactions (Visa and Mastercard). Our model considers the three roles present 
in a transaction, and supports the fine-grained analysis of all relevant 
security properties.

Second, we identify and demonstrate, for the first time in actual terminals, 
a practical attack that allows attackers to make high-value payments with the 
victim's card, without knowledge of the card's PIN. We also identify an attack 
that allows one to effectively steal goods by tricking terminals into accepting 
unauthentic offline transactions. Our attacks demonstrate that EMV's liability 
shift should be voided because credit card fraud is not necessarily the result 
of negligent behavior of consumers or merchants.

Finally, based on our full-scale, automatic, Tamarin-supported analysis of 
EMV's fundamental security properties, we identify the EMV configurations that 
guarantee secure transactions. Based on these configurations, we propose 
solutions that can be implemented in the payment terminals and rule out 
security breaches.

Note that our focus is on EMV's design, not implementations themselves. In this 
way, we can end the penetrate-and-patch arms race where attackers continually 
find and exploit protocol weaknesses. Of course this is only one part of the 
overall picture, as attackers can still exploit implementation weaknesses; but 
it is a substantial part and is also a prerequisite for any ``full stack'' 
effort to formally develop a verified protocol down to the level of code.

\subsection*{Organization}

In Section~\ref{sec:related-work} we describe related work, focusing on 
previous EMV security analyses. In Section~\ref{sec:emv-descrition} we provide 
background on the EMV protocol. In Section~\ref{sec:symbolic-model} we 
present our formal model of EMV, focusing on how we model EMV's numerous  
configurations and how we define and analyze its security properties. In 
Section~\ref{sec:results} we present the results of this analysis. Later, in 
Section~\ref{sec:attack-and-fixes}, we describe an Android app that 
we developed and used to show that our Tamarin findings can be turned into 
real-world attacks. We also suggest improvements to 
terminals that guarantee secure transactions. We draw conclusions in 
Section~\ref{sec:conclusion}.

\subsection*{Ethical Considerations}
We carried out all our tests using our own credit and debit cards. On April 
30th, 2020, we notified Visa of the attacks discovered. Also, we will not make 
our Android app available to the public, at least not until the reported issues 
are resolved.

\section{Related Work}\label{sec:related-work}

Given its financial importance, it is not surprising that the EMV standard has 
been extensively studied. We review here the most relevant related work. This 
previous work concerns either implementation flaws or protocol flaws 
discovered by analyzing selected and possibly simplified parts of the EMV 
specification. In contrast, our analysis integrates all the different 
configurations for card authentication, cardholder verification, and 
transaction authorization in a single symbolic model. This provides a basis not 
only for finding all relevant design errors, but also producing correctness 
proofs.

In 2010, Murdoch \etal{}~\cite{MurdochDAB10} identified a serious flaw in EMV's 
offline \acrlong{CVM}s (\acrshort{CVM}s). Namely, the card's response to the 
terminal's PIN verification request is not authenticated. Therefore, a 
man-in-the-middle (MITM) could reply with the \emph{success} message to 
\emph{any} PIN the terminal would request verification for. The dummy PIN could 
be blocked from reaching the card, which would then assume that either the 
chosen \acrshort{CVM} was paper signature or no \acrshort{CVM} was required at 
all. All subsequent steps would be carried out normally and the transaction 
would be accepted.

Even though Murdoch \etal{}'s attack comes with some engineering challenges, 
such as miniaturizing the MITM infrastructure, these challenges 
appear to have been overcome as observed in the aforementioned forgery of 
credit cards in France and Belgium~\cite{FerradiGNT16}. 
Our analysis demonstrates that this attack still exists in old cards that 
support neither asymmetric cryptography nor online PIN verification (see 
Section~\ref{sec:results-contact}). Unfortunately, 
many modern cards that support both features are still vulnerable to our own 
PIN bypass attack, which we present in this paper.

Soon after, De Ruiter and Poll~\cite{RuiterP11} gave a 
ProVerif~\cite{proverif01} model of a variant of the EMV contact protocol. They 
summarize over 700 pages of EMV specifications into 370 lines of F\# 
code, which they transform into the ProVerif language using the FS2PV 
tool~\cite{BhargavanFGT06}. Their analysis misses the attack 
of~\cite{MurdochDAB10} because the terminal's 
selection of the \acrshort{CVM} is over-simplified to always opt for the 
offline plaintext PIN. This makes the card always expect a PIN verification 
request, with the correct PIN, before continuing with the transaction. 

Some of EMV's flaws have also been identified from empirical 
studies in the field~\cite{RolandL13,BondCMSA14,EmmsAFHM14}. For 
example, Bond \etal{}~\cite{BondCMSA14}, together with unsatisfied consumers 
who were denied refunds for fraud claims, were given access to the bank 
logs of the disputed transactions. This access, together with 
reverse-engineering some ATMs, revealed flawed implementations of EMV. They 
noted that the supposedly unpredictable numbers generated by some terminals 
were actually pretty predictable, allowing criminals to \emph{pre-play} 
payments and use the retrieved data for later purchases.

Bond \etal{} also report that a pre-play attack is still 
possible even when the terminal's random number generator works correctly. In 
this case, the pre-play consists of the attacker replacing the 
terminal-generated nonce with one used in an earlier transaction between the 
attacker and the victim's card. 

Symbolic models consider the Dolev-Yao threat model~\cite{DolevY83}, where the 
adversary only knows public knowledge, the data sent over the network, and the 
outcome of public functions on known input. The adversary is also an active 
attacker, who can modify, block, and inject data on the network. In 
this model, however, random number generators are assumed to be sound, i.e., 
random numbers cannot be predicted. Therefore, attacks of this kind are usually 
not part of a symbolic analysis that examines the specification (not the 
implementation) for logical errors. Our analysis thus does not uncover Bond 
\etal{}'s attacks~\cite{BondCMSA14}. Note that it is possible though to 
incorporate weak random generators and compromised channels into symbolic 
models, as described in~\cite{BasinC14}.



The EMV contactless protocol's security has been challenged multiple 
times too. For example, Roland and Langer~\cite{RolandL13} detected a 
downgrade attack that exploits Mastercard's \emph{MagStripe} mode, a legacy 
authentication mode kept for backward compatibility. They showed that a mobile 
phone supporting \acrfull{NFC} can collect all authentication codes that a 
card could produce in response to all potential challenges from a terminal. 
Hence, a clone card pre-loaded with the codes can be used for fraudulent 
payments. This attack is feasible because the MagStripe mode 
reduces the terminal's pool of unpredictable numbers to 1000 values only. 
In this paper we do not consider the MagStripe mode because the random 
generators are assumed sound (as explained above) and this mode has been 
deprecated in many countries.

Other attacks demonstrated against EMV contactless payment protocols are 
well-known \emph{relay 
attacks}~\cite{DrimerM07,FrancisHMM11,SportielloC13,ChothiaGRBT15,BocekKTS16}. 
The works~\cite{DrimerM07,ChothiaGRBT15,BocekKTS16} suggest using distance 
bounding protocols~\cite{BethD90,BrandsC93} as a countermeasure to such 
attacks. Although distance bounding does prevent relay attacks, only 
Mastercard seems to be inclined to use it. Relay attacks are usually ignored 
because they are presumably feasible only for small transactions, since larger 
transactions require cardholder verification.


In 2014, Emms \etal{}~\cite{EmmsAFHM14} observed that some 
UK-issued contactless Visa credit cards drop the PIN verification for  
transactions in foreign currencies. The authors developed a proof-of-concept 
implementation of the attack, where they faked a transaction of almost one 
million US dollars. We attempted to reproduce the experiments 
of~\cite{EmmsAFHM14} but all 
modern cards we tested did ask for PIN verification for high-value 
transactions in both domestic and foreign currencies. 


There exist various symbolic models that showcase the EMV contactless 
protocols~\cite{MauwSTT18,ChothiaRS18,DebantDW18,MauwSTT19,DebantD19}.
 All of these focus on verifying proximity between the card and the terminal. 
 They also consider the terminal and the bank as a single agent and 
 consequently do not cover the pre-play attack of~\cite{RolandL13}. %
%

Galloway and Yunusov~\cite{GallowayY19} recently presented a 
man-in-the-middle attack that also circumvents Visa's PIN verification. 
Their attack is similar to ours in that it modifies a card-sourced message that 
instructs the terminal that cardholder verification was performed on the 
consumer's device. In contrast to our attack, Galloway and Yunusov's attack 
also modifies a terminal-sourced message in which the cardholder verification 
request is encoded. According to EMV's (generic) cryptogram definition, such 
message should be protected against modification. Their attack works because 
Visa's proprietary cryptograms do not prevent such modification, or at least 
not the ones implemented by the cards they tested. Interestingly, and 
worrisome, our own attack demonstrates that the \emph{strongest} cryptogram 
proposed by EMV still does not suffice to correctly verify the cardholder. 
The details are given in Section~\ref{sec:attack-and-fixes}.

\section{EMV Description}\label{sec:emv-descrition}

The EMV specification runs over 2,000 pages split across several books. 
Moreover, many of the statements in these books are quite complex and 
cross-reference other books. In this section we give a detailed description of 
the standard. Given its complexity, creating this specification and its formal 
model in Tamarin was a major undertaking that took over six months of 
full-time work. Our methodology included not only carefully reading the 
standard, but also cross-checking and disambiguating 
%
its statements with data from over 30 real-world transaction logs that we 
obtained using the Android app we developed, described in later 
sections.

An EMV transaction consists of a series of \acrfull{APDU} command/response 
exchanges and can be divided into four phases:
\begin{enumerate}[label={\itshape\Alph{enumi}.}]
\item \emph{Initialization}: the card and the terminal agree on the 
application to be used for the transaction and exchange static data 
such as the card's records containing information about the card and the 
issuing bank (or simply the bank from now on, unless otherwise specified).
\item \emph{\acrfull{ODA}}: the terminal performs a 
PKI-based validation of the card. Once the card has provided the 
terminal with the \acrfull{CA} index, the bank's \acrfull{PK} 
certificate issued by the \acrshort{CA}, and the card's \acrshort{PK} 
certificate issued 
by the bank, the terminal validates the card's signature on the transaction 
details.

\item \emph{Cardholder Verification}: the terminal determines whether 
the person presenting the card is the legitimate cardholder. This is done 
using a method that the card and the terminal both support. The most common 
method is online enciphered PIN verification, in which the terminal sends (an 
encryption of) the entered PIN to the bank for verification. The card is not 
involved.
\item \emph{\acrfull{TA}}: the transaction is declined 
offline, accepted offline, or sent to the issuing bank for online authorization.
\end{enumerate}

\begin{figure*}
\centering
\includegraphics[scale=.55]{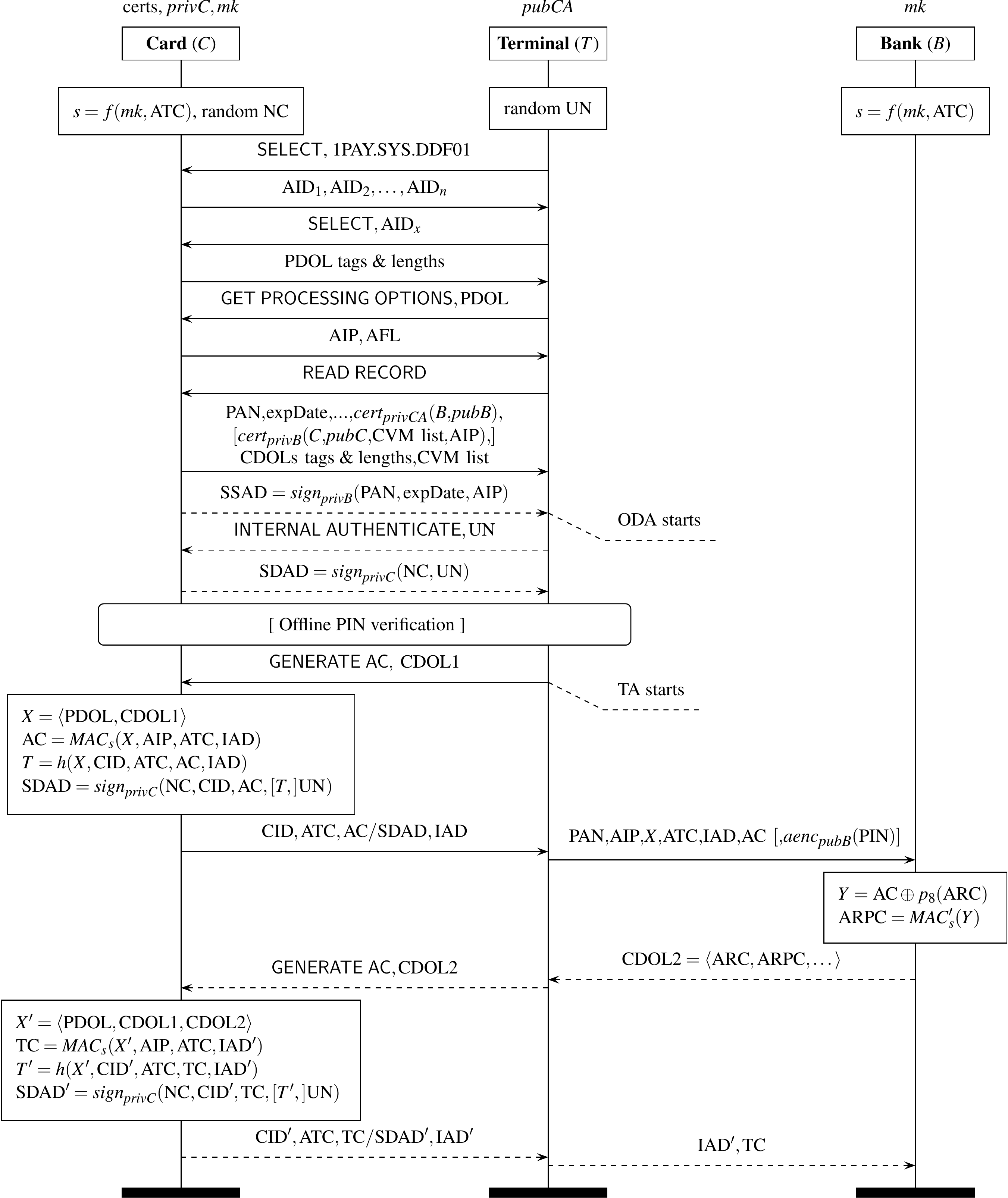}
\caption{An overview of the EMV transaction. Dashed 
messages and bracketed terms are either 
optional, or depend on previous steps, or depend on the parties' choices. 
For simplicity, this chart only shows the execution flow in which 
the card responses have the success trailer \hexfont{9000}. %
\hspace{\textwidth}
Notation: $\oplus$ is exclusive-OR; $f$ 
is a key derivation function; $(\skCard,\pkCard)$, $(\skBank,\pkBank)$, and 
$(\skCA,\pkCA)$ are the PKI pairs of the card, the bank, and the \acrshort{CA}, 
respectively; $\cert{cont}{k}$ is the PKI certificate on $cont$ 
signed with the private key $k$; $\sign{m}{k}$ is the signature on $m$ with the 
key $k$; $\aenc{m}{k}$ is the asymmetric encryption of $m$ with the key $k$; 
$\MAC{m}{k}$ and $\MACprime{m}{k}$ are \acrlong{MAC}s on $m$ 
with the key $k$; $p_b(m)$ is the 
right-padding of $m$ with $b$ zero bytes.
}\label{fig:emv-full}
\end{figure*}

An overview of the full EMV transaction flow is depicted in 
Figure~\ref{fig:emv-full} and the details of each phase are given next.

\subsection{Initialization}\label{sec:initialization}

The first step of an EMV transaction is the 
application selection. The terminal issues the \cmdfont{SELECT} command with 
the string \datafont{1PAY.SYS.DDF01} (in bytes), which refers to the 
contact \acrfull{PSE}, or \datafont{2PAY.SYS.DDF01} for 
contactless. The card responds with the sequence of \acrlong{AID}s 
(\acrshort{AID}s)
. In the response, the card may also request the \acrfull{PDOL}, which is a 
list of terminal-sourced transaction data. The \acrshort{PDOL} typically 
includes the amount, the country code, the currency, the date, the 
transaction type, and the terminal's random number \acrshort{UN} (called 
\acrlong{UN} in EMV's terminology).

The terminal issues the \cmdfont{GET~PROCESSING~OPTIONS} command along with 
the \acrshort{PDOL} data, if requested by the card. The card responds with the 
2-byte \acrlong{AIP} (\acrshort{AIP}, which indicates the 
supported authentication methods and whether cardholder verification is 
supported) and the \acrlong{AFL} (\acrshort{AFL}, which points to a list of 
files and records 
that the terminal should read from the card). The terminal then learns these 
records using the \cmdfont{READ~RECORD} command. The records typically include:
\begin{itemize}
\item the \acrlong{PAN} (\acrshort{PAN}, commonly known as the card 
number), the card's expiration date, and other static data;
\item the index of the \acrshort{CA}, the bank's \acrshort{PK} certificate 
issued 
by the \acrshort{CA}, and the card's \acrshort{PK} certificate issued by the 
bank, 
if the card supports asymmetric encryption;
\item the first and second \acrlong{CDOL}s (\acrshort{CDOL}1 and 
\acrshort{CDOL}2, respectively), which typically include the \acrshort{PDOL} 
and further transaction data; and\item the list of the supported 
\acrshort{CVM}s.
\end{itemize}

From the \acrshort{CA}'s index, the terminal retrieves the \acrshort{CA}'s 
\acrshort{PK} from an 
internal data base and then verifies the bank's certificate. Afterwards, from 
the bank's certificate, the terminal acquires the bank's \acrshort{PK} and 
verifies the card's certificate, if applicable. Finally, the terminal acquires 
the card's \acrshort{PK} from the card's certificate.

\subsection{Offline Data Authentication}\label{sec:oda}
There are three methods for \acrfull{ODA}, also known as Card Authentication:

\begin{enumerate}
\item \emph{\acrfull{SDA}}: the card supplies the 
terminal with the \acrfull{SSAD}, which is the 
bank's signature on the card's static data such as 
the \acrshort{PAN}, the card's expiration date, and 
optionally the \acrshort{AIP}. The \acrshort{SDA} method prevents modification 
of the card's static data, but it does not prevent cloning.

\item \emph{\acrfull{DDA}}: the terminal transmits 
the \cmdfont{INTERNAL~AUTHENTICATE} command whose payload is the 
\acrfull{DDOL}. The \acrshort{DDOL} must contain the terminal's \acrlong{UN}. 
In response to this challenge, the card transmits 
the \acrfull{SDAD}, which is the card's signature 
on the card's dynamic data (a fresh number $\nc$) and the received 
\acrshort{DDOL}. The \acrshort{DDA} method protects against modification of 
card data and cloning.

\item \emph{\acrfull{CDA}}: this is similar to 
\acrshort{DDA} but it includes the transaction details in the \acrshort{SDAD}, 
e.g., the transaction amount. 
\end{enumerate}

\subsection{Cardholder Verification}
A \acrfull{CVM} can be a paper signature, PIN verification, \acrfull{CDCVM}, or 
a combination of these. There are three specific methods for PIN verification: 
\begin{enumerate}
\item \emph{Offline Plaintext PIN} (or simply \emph{plain PIN}): the 
terminal sends the \cmdfont{VERIFY} command along 
with the entered PIN and the card responds with the success message 
\hexfont{9000} if the PIN is correct, or the 
failure message $\hexfont{63C}x$, where the digit $x$ is the number of 
tries left. When no tries remain, i.e., $x=0$, then 
the card must respond with the PIN-blocked message \hexfont{6983} to any 
subsequent \cmdfont{VERIFY} requests.

\item \emph{Offline Enciphered PIN} (or simply \emph{enciphered PIN}): the 
terminal sends the \cmdfont{GET~CHALLENGE} command and the card responds with a 
random number. 
Then the terminal issues the \cmdfont{VERIFY} command whose payload is 
the encryption, with the card's \acrshort{PK}, of the entered PIN, the received 
random number, and random padding generated by the terminal. Upon reception, 
the card decrypts the payload and responds accordingly, using the messages 
described in the plain PIN method.

\item \emph{Online Enciphered PIN} (or simply \emph{online PIN}): the card is 
not involved. Instead, the terminal sends the entered PIN encrypted to the 
issuing bank when requesting the transaction authorization.
\end{enumerate}

The \acrlong{CDCVM} is intended to be performed by devices such as mobile 
phones, which authenticate the cardholder through fingerprint or face 
recognition. How the terminal and the device conduct \acrshort{CDCVM} is out of 
EMV's scope. Nevertheless, this method is the fundamental cause of one of the 
new attacks that we report on in this paper.

\subsection{Transaction Authorization}

The terminal can decide either to decline the transaction offline, to authorize 
the transaction offline, or to request online authorization from 
the issuing bank. This decision is made based on various checks such as 
the offline ceiling limit, above which transactions should be processed online.

The terminal sends the \cmdfont{GENERATE~AC} command to the card, 
along with the \acrshort{CDOL}1. This command instructs 
the card to supply the 8-byte \acrfull{AC} which is:
\begin{itemize}
\item an \acrfull{AAC}, if the terminal 
decided to decline the transaction,
\item a \acrfull{TC}, if the terminal decided to 
approve the transaction offline, or
\item an \acrfull{ARQC}, if the terminal 
decided to request online authorization.
\end{itemize}

The requested type of cryptogram is encoded in the command payload. The card 
then issues the \acrshort{AC} whose type can be either the requested one, or an 
\acrshort{ARQC}, or an \acrshort{AAC}. The card 
cannot generate a \acrshort{TC} when an \acrshort{ARQC} was requested. 

The cryptogram is a block cipher-based \acrfull{MAC} computed over the 
transaction 
details, the \acrshort{AIP}, and the \acrlong{ATC} (\acrshort{ATC}, which is a 
2-byte counter incremented on every transaction). The 
\acrshort{MAC}'s key is a 
session key $\Ks$ derived from the \acrshort{ATC} and a symmetric master key 
$\Km$ shared by the bank and the card. Along with the cryptogram itself, the 
card sends other data such as the 1-byte \acrlong{CID} (\acrshort{CID}, which 
indicates the type of cryptogram being sent), the transaction counter 
\acrshort{ATC}, and if \acrshort{CDA} was requested in the command payload, the 
\acrfull{SDAD} replaces the \acrshort{AC}. In this case, the \acrshort{SDAD} is 
a 
signature on the 
card's random number $\nc$, the \acrshort{CID}, the cryptogram, a hash of the 
transaction details, and the terminal's \acrshort{UN}. 

If the card responds with a \acrshort{TC} and the chosen \acrshort{CVM} was not 
online PIN, then the transaction is approved and the \acrshort{TC} serves as a 
settlement to instruct the bank to transfer the funds to the merchant's 
account. 

If the transaction must be authorized online, then the terminal forwards to the 
bank the transaction details, the \acrshort{ARQC}, and if online 
PIN verification was the selected \acrshort{CVM}, then also the entered PIN. 
The bank authorizes or declines the transaction by sending back to the terminal 
the 2-byte \acrlong{ARC} (\acrshort{ARC}, authorize/decline and further data) 
and the \acrfull{ARPC}. The latter is also a cipher-based \acrshort{MAC} 
generated over the exclusive-OR of the \acrshort{ARC} (padded to 8 bytes) and 
the received cryptogram \acrshort{ARQC}, using the session key $\Ks$. The 
terminal then issues the \cmdfont{EXTERNAL~AUTHENTICATE} command (or 
equivalently a second \cmdfont{GENERATE~AC}) to inform the card of the bank's 
decision. The card constructs the response analogously to its response to the 
(first) \cmdfont{GENERATE~AC} command, only this time no \acrshort{ARQC} is 
sent, but instead either a \acrshort{TC} or an \acrshort{AAC}.

\section{Modeling and Analysis Methodology}\label{sec:symbolic-model}

To model and analyze the EMV standard, we use the protocol verification tool 
Tamarin~\cite{tamarin13, SchmidtMCB12}. Tamarin is a state-of-the-art 
model-checker for security protocol verification. It features an expressive 
language for specifying protocols, their properties, and adversaries, as well 
as powerful inference procedures for automating much of protocol verification. 
We first provide some background on Tamarin and then present the properties we 
analyze and our analysis methodology.

\subsection{Tamarin Background}
In Tamarin's underlying theory, cryptographic messages are terms in an 
order-sorted term algebra $(\sort, \leq, \term)$ where $\sort$ is a set of 
sorts, $\leq$ is a partial order on $\sort$, $\signature$ is a signature, and 
$\variables$ is a countably infinite set of variables. 
%
For example, the term $\pubkey{k}$, with $\mathit{pk}\in\signature$, denotes 
the public key associated to the private key $k\in\term$. Similarly, the term 
$\aenc{m}{k}$, with $\mathit{aenc}\in\signature$, denotes the asymmetric 
encryption of the message $m\in\term$ with the public key $k\in\term$. The 
algebraic properties of the cryptographic functions are defined by equations 
over terms. For example, $\adec{\aenc{m}{\pubkey{k}}}{k} = m$ specifies the 
semantics of asymmetric decryption.

Tamarin models a protocol's set of executions as a labeled transition system 
(LTS). The states of the LTS are multisets of \emph{facts}, which formalize 
the local states of the agents running the protocol, the adversary's knowledge, 
and messages on the network. Facts are of the form 
$\fact{F}(a_1,a_2\dots,a_n)$ where $\fact{F}$ is a symbol 
from an unsorted signature $\Gamma$ of predicate symbols and 
$a_i\in\mathcal{T}_{\Sigma}(\mathcal{V})$. 
%
Transitions between states are determined by \emph{transition rules} (or simply 
rules). A rule is a triple $(l,a,r)$, also written as $\makerule{l}{a}{r}$, 
where $l$, $a$, and $r$ are multisets of facts. For example, the following rule 
specifies the transmission of the hash of a received message:
\begin{align*}
\makerule{\fact{In}(m)}{\fact{SentHash}(A, m)}{\fact{State1}(A,m), 
\fact{Out}(h(m))}\text{.}
\end{align*}
This rule states that, if there is a term $m$ input on the network, then update 
the local state of $A$ to $\fact{State1}(A,m)$, remove $m$ from the network, 
and output the term $h(m)$ on the network, possibly for reception by $A$'s 
communication partner. The transition is labeled with $\fact{SentHash}(A, m)$, 
meaning that $A$ sent the hash of $m$.

In what follows, let $\facts$ be the universe of 
facts and $\rules$ the universe of rules. Whereas $\powerset{.}$ 
denotes the power set of a set, we use $\powermultiset{.}$ to refer to the 
power \emph{multiset} of a set. We define 
the function $linear\colon\powermultiset{\facts}\to\powermultiset{\facts}$ that 
yields all \emph{linear} facts from the input multiset of facts. Linear facts 
model resources that can be consumed just once, such as messages on the 
network. Facts that are not linear are called \emph{persistent} and can be 
reused arbitrarily often without being consumed. We also define the function 
$gins\colon \powerset{\rules}\to \powerset{\rules}$ that yields the set of all 
\emph{ground instances} of the input set of rules. A ground instance of a rule 
is the rule resulting from the substitution of all variables with \emph{ground 
terms} (i.e., terms from $\mathcal{T}_{\signature}$). Also, let 
$\specialrules\subseteq\rules$ be the set of global rules modeling a network 
controlled by a Dolev-Yao adversary~\cite{DolevY83} as well as the generation 
of random, fresh values.

A protocol $P\subseteq\rules$ is a set of rules. The associated LTS is $(S, 
\Lambda, \xrightarrow{})$, where $S = \powermultiset{\facts}$, $\Lambda = 
\powermultiset{\facts}$, and $\xrightarrow{}\ \subseteq S \times \Lambda\times 
S$ is defined by:
\begin{align*}
s\xrightarrow{a}s' \iff&\exists (l,a,r)\in gins(P\cup \specialrules).\\ 
&\hspace{2ex}l\subseteq s \wedge s' = ( s\setminus linear(l)) \cup r\text{.}
\end{align*}

A transition consumes the linear facts of $l$ from the current 
state, adds the facts from $r$, and labels the transition with $a$. An 
execution of $P$ is a finite sequence $(s_0, a_1, s_1,\ldots, a_n, s_n)$ such 
that $s_0 = \emptyset$ and $s_{i-1}\xrightarrow{{a_i}}s_i$ for 
all $1\leq i\leq n$. The sequence $(a_1,\dots,a_n)$ is a \emph{trace} of $P$ 
and the set of all of $P$'s traces is denoted $\tracesOf{P}$. Security 
properties are specified using first-order logic formulas on traces. Further 
details on Tamarin's syntax and semantics can be found 
in~\cite{tamarin13,SchmidtMCB12}.

\subsection{Security Properties}

As we have seen, EMV involves three parties: the consumer's card, the 
merchant's terminal, and the cardholder's bank. Its central security properties 
concern the parties authenticating each other, guarantees on transaction 
information, and the secrecy of sensitive data. 

The first property we examine is that no 
terminal-accepted transaction will be declined by the bank. This property is 
particularly relevant for offline-capable terminals, which typically do not 
request online authorization for low-value transactions. Such terminals can 
be cheated if the property fails.

\begin{definition}[Bank accepts]\label{def:bank-accepts}
A protocol $P$ satisfies the property that the \emph{bank accepts 
terminal-accepted transactions} if for every $\trace\in\tracesOf{P}$:
\begin{align*}
\forall \transaction, i.\ &\fact{TerminalAccepts}(\transaction)\in 
\trace_i\implies\\
&\hspace{3ex}\nexists j.\ \fact{BankDeclines}(\transaction)\in \trace_j\ 
\vee\\
&\hspace{3ex}\exists A, k.\ \fact{Honest}(A)\in \trace_i \wedge 
\fact{Compromise}(A)\in 
\trace_k\text{.}
\end{align*}

\end{definition}

In our model, the $\fact{TerminalAccepts}(\transaction)$ fact is added to the 
trace only if the terminal is satisfied with the transaction $\transaction$ 
and the associated cryptographic proofs provided by the card. That is, when the 
terminal issues a purchase receipt. The $\fact{BankDeclines}(\transaction)$ 
fact is produced when the bank receives an authorization request for the 
transaction with a wrong Application Cryptogram. The last line rules out 
transactions where an agent, presumed honest, has been compromised. For 
example, a bank that maliciously rejects a correct transaction should not make 
the property fail.

Our second property corresponds to the authentication 
property commonly known as \emph{injective agreement}~\cite{Lowe97a,CremersM12}.


\begin{definition}[Authentication to terminal]\label{def:auth-to-terminal}
A protocol $P$ satisfies \emph{authentication to the terminal} if for every 
$\trace\in\tracesOf{P}$:
\begin{align*}
&\forall T, P, r, \transaction, i.\\
&\hspace{1ex}\fact{Commit}(T, P, \left\langle r, 
\literalstr{Terminal}, \transaction\right\rangle )\in 
\trace_i\implies\notag\\
&\hspace{2ex}\big(\exists j.\ \fact{Running}(P, T, \left\langle r, 
\literalstr{Terminal}, \transaction\right\rangle )\in \trace_j\ \wedge \notag\\ 
&\hspace{3ex}\nexists i_2, T_2,P_2.\notag\\
&\hspace{4.5ex}\fact{Commit}(P_2,T_2, \left\langle r, 
\literalstr{Terminal}, \transaction\right\rangle)\in \trace_{i_2}\wedge i_2\neq 
i\big)\ 
\vee\notag\\
&\hspace{3ex}\exists A, k.\ \fact{Honest}(A)\in \trace_i \wedge 
\fact{Compromise}(A)\in 
\trace_k\text{.} \notag
\end{align*}
\end{definition}

The above property, with 
$\literalstr{Terminal}\in\groundterm$ and $\left\langle \right\rangle 
\in\signature$, states that whenever the 
terminal $T$ \emph{commits} to a 
transaction $\transaction$ with its communication partner $P$, then 
either $P$, in role $r\in 
\{\literalstr{Card},\literalstr{Bank}\}\subseteq\groundterm$, was 
\emph{running} the protocol with $T$ and they agree on $\transaction$, or an 
agent, presumed honest, has been compromised. Additionally, there is a unique 
\fact{Commit} fact for each pair of accepted transaction and accepting 
agent, which means that replay attacks are prevented.

The facts \fact{Commit} and \fact{Running}, introduced in~\cite{Lowe97a}, are 
used to specify authentication properties. A \fact{Commit} fact represents an 
agent's belief about its communication partner's local state, whereas 
\fact{Running} represents the partner's actual state. Authentication 
properties are therefore expressed in terms of matching pairs of such facts. 
In our models, \fact{Commit} facts occur whenever the committing agent is in a 
satisfactory state when the transaction is ready to be accepted. 

Our third property is also an authentication property and is very
similar to the 
second, except that the agent 
who commits is the bank. 
That is, the 
definition is the same except the ground 
term $\literalstr{Terminal}$ is now $\literalstr{Bank}\in \groundterm$. %
%

Another property relevant for formal protocol 
analysis is \emph{secrecy} (a.k.a. confidentiality). The secrecy of a term $x$ 
holds when $x$ is not known to the attacker. The attacker's knowledge of a 
term $x$ is written as $\fact{KU}(x)$, where $\fact{KU} \in \Gamma$ is a fact 
symbol defined by Tamarin's built-in rules that model how the attacker acquires 
knowledge. The definition of secrecy also assumes that the 
agents involved are not compromised.

\begin{definition}[Secrecy]\label{def:secrecy}
A protocol $P$ satisfies \emph{secrecy} if for every $\trace\in 
\tracesOf{P}$:
\begin{align*}
\forall x,i.\ &\fact{Secret}(x)\in \trace_i \implies\\ 
&\hspace{2ex}\nexists j.\ \fact{KU}(x)\in \trace_j\;\vee\\
&\hspace{2ex}\exists A, k.\ \fact{Honest}(A)\in \trace_i \wedge 
\fact{Compromise}(A)\in \trace_k\text{.}
\end{align*}
\end{definition}

In an EMV transaction, terms that should be secret include the 
PIN number, the \acrshort{PAN} (i.e., the card number), and the keys (i.e., 
private keys and symmetric shared keys). 

We also consider other properties such as \emph{executability}, 
which allows one to assess whether a protocol execution reaches a state where 
the bank and the terminal have accepted a transaction and no compromises have 
occurred. This represents a sanity check showing that the protocol modeled 
behaves as expected and allows the executions of protocol runs without 
adversary involvement. This ensures that there are no modeling errors that 
would make the specified protocol inoperable and lead to false results.

\begin{definition}[Executability]\label{def:exec}
A protocol $P$ is \emph{executable} if $\trace\in\tracesOf{P}$ exists such 
that:
\begin{align*}
&\exists \transaction,C,B,T,i,j,k,l.\\
&\hspace{2ex}\fact{Running}(C, T, \left\langle \literalstr{Card}, 
\literalstr{Terminal}, \transaction\right\rangle )\in \trace_i\ \wedge\\
&\hspace{2ex}\fact{Commit}(T, C, \left\langle \literalstr{Card}, 
\literalstr{Terminal}, \transaction\right\rangle )\in \trace_j\ \wedge\\
&\hspace{2ex}\fact{Running}(C, B, \left\langle \literalstr{Card}, 
\literalstr{Bank}, \transaction\right\rangle )\in \trace_k\ \wedge\\
&\hspace{2ex}\fact{Commit}(B, C, \left\langle \literalstr{Card}, 
\literalstr{Bank}, \transaction\right\rangle )\in \trace_l\ \wedge\\
&\hspace{2ex}\nexists A, a.\ \fact{Compromise}(A)\in \trace_a\text{.}
\end{align*}
\end{definition}

\subsection{Analysis Methodology}\label{sec:methodology}

We construct our model in a way that accounts for all possible protocol 
executions and interactions, but gives us a structured analysis of which kinds 
of executions are vulnerable to attacks. We start by formalizing the EMV 
standard in two \emph{generic} models: 
\begin{enumerate}
\item one for the {\bfseries EMV contact protocol}, modeling the full 
execution space of a contact transaction, and
\item one for the {\bfseries EMV contactless protocol}, modeling the full 
execution space of a Mastercard~\cite{mastercard19} or Visa~\cite{visa19} 
contactless transaction.
\end{enumerate}

Each of these two models captures all possible executions of the corresponding 
\acrlong{PSE} (contact or contactless), including simultaneous 
transactions with different cards, terminals, types of authentication, 
cardholder verification methods, and all the other settings. For 
example, the contactless protocol model allows for executions between a 
terminal, which believes to be in a Visa transaction, and three cards, which 
may be different from Visa cards. Clearly, whether the system can 
reach a state where the transaction is accepted depends on the actual messages 
and cryptographic proofs that the terminal and the bank receive.

Tamarin exhibits a property violation by constructing a trace that contradicts 
the given property. Clearly, Tamarin cannot output all such traces 
as there are infinitely many (simply by adding unrelated steps), if one exists. 
Running Tamarin on the generic models will therefore either lead to a 
successful verification or one attack trace, violating the property, with the 
``least secure'' type of card and authentication method, among other settings. 
However, one might be interested, for example, in the property of 
authentication to the bank specifically for transactions where the card used 
\acrlong{CDA} (\acrshort{CDA}, recall from Section~\ref{sec:oda}) and the 
transaction value was high, i.e., above the \acrshort{CVM}-required limit.

With this in mind, we employed a modeling strategy that automatically generates 
specific Tamarin models from the two generic models. To automatically generate 
the specific models, we use \emph{target configurations}. A target 
configuration is a choice of arguments that selects the transactions for which 
we want to verify the security properties. A generic model and a target 
configuration determine what we call a \emph{target model}. For example, 
Visa\_DDA\_Low is a target model generated from the contactless protocol 
(generic) model with the target arguments:
\begin{itemize}
\item \acrshort{DDA}: referring to the offline data authentication 
method (known as \emph{fast} \acrshort{DDA} in~\cite{visa19}), and
\item Low: indicating a low-value transaction.
\end{itemize}

We automated the generation of target models and the interested reader can find 
the technical details in Appendix~\ref{app:target-models-generation} as well as 
in our Tamarin theories and their README~\cite{repo}.

In our models, we consider the following transaction data to be agreed upon for 
the authentication properties (i.e., the 
term $\transaction$ in Definition~\ref{def:auth-to-terminal}):
\begin{itemize}
\item the \acrfull{PAN};
\item the \acrfull{AIP};
\item the \acrfull{CVM} used;
\item the \acrfull{ATC};
\item the \acrfull{AC} data input ($X$ and $X'$ in Figure~\ref{fig:emv-full});
\item the \acrfull{AC} itself; and
\item the \acrfull{IAD}.
\end{itemize}

For both the contact and contactless models, between the terminal and the card 
(and vice versa) we modeled a channel controlled by the Dolev-Yao 
adversary, who can listen, block, inject, and modify the 
transmitted data. Between the bank and the terminal (and vice versa) we modeled 
a secure channel that offers authentication and secrecy.


We also assumed that terminals do not complete high-value, contactless 
transactions with cards that (apparently) do not support cardholder 
verification. In such transactions, the common practice of 
terminals is to reject the attempt and instruct the cardholder to switch to the 
contact interface.

\section{Analysis Results}\label{sec:results}

We conducted a full-scale, automated security analysis of 40 configurations 
of EMV, including both types of transactions: contact and contactless. 
We describe the results of this comprehensive analysis in this section.

\subsection{Analysis Results for the EMV Contact Protocol}
\label{sec:results-contact}

Our analysis results for the 24 configurations of the EMV contact protocol are 
summarized in Table~\ref{tab:results-contact}. Although there are no major 
surprises here, the results illustrate the benefits of a comprehensive 
formalization and analysis. In particular, we both rediscovered existing, 
known attacks on the contact protocols as well as attacks that, to our 
knowledge are new, but relatively difficult to carry out in practice and 
therefore have limited practical relevance. Note that we have omitted the 
results for secrecy from the table because they are identical for all models. 
All of our models and proofs are available at~\cite{repo}.

\begin{table*}
\caption{Analysis results for the EMV contact protocol. All target models have 
55 rules. The last two columns indicate, in that order, %
the number of lines of Tamarin code that the model comprises, and the time 
taken for our analysis, using Tamarin v1.5.1 on a computing server 
running Ubuntu 16.04.3 with two Intel(R) Xeon(R) E5-2650 v4 @ 2.20GHz CPUs 
(with 12 cores each) and 256GB of RAM. Here we used 10 threads and at most 20GB 
of RAM per model. 
The models for which all four properties were verified are highlighted in 
bold.
}
\label{tab:results-contact}
\renewcommand{\arraystretch}{1.35}
\centering
\begin{tabular}{l l c c c c c r}
\hline\hline
\multirow{2}{*}{\bfseries No.} & 
\multirow{2}{*}{\bfseries Target model} & 
\multicolumn{4}{c}{\bfseries Properties}
& \multirow{2}{*}{\bfseries LoC}
& \multirow{2}{*}{\bfseries Time}
\\\cline{3-6}
& & \bfseries \itshape executable & \bfseries \itshape bank accepts & 
\bfseries \itshape auth. to terminal & \bfseries \itshape auth. to bank & &
\\\hline
1 & Contact\_SDA\_PlainPIN\_Online & \verified{} & \falsified{}$^{(2)}$ & 
\falsified{}$^{(1,2)}$ & \falsified{}$^{(1)}$ & 758 & 13m07s\\
2 & Contact\_SDA\_PlainPIN\_Offline & \verified{} & \falsified{}$^{(2)}$ & 
\falsified{}$^{(1,2)}$ & \falsified{}$^{(1)}$ & 761 & 11m39s\\
3 & Contact\_SDA\_OnlinePIN\_Online & \verified{} & \falsified{}$^{(2)}$ & 
\falsified{}$^{(1,2)}$ & \falsified{}$^{(1)}$ & 758 & 13m02s\\
4 & Contact\_SDA\_OnlinePIN\_Offline & -- & -- & -- & -- & 731 & 11m48s\\
5 & Contact\_SDA\_NoPIN\_Online & \verified{} & \falsified{}$^{(2)}$ & 
\falsified{}$^{(1,2)}$ & \falsified{}$^{(1)}$ & 752 & 8m21s\\
6 & Contact\_SDA\_NoPIN\_Offline & \verified{} & \falsified{}$^{(2)}$ & 
\falsified{}$^{(1,2)}$ & \falsified{}$^{(1)}$ & 755 & 6m37s\\
7 & Contact\_SDA\_EncPIN\_Online & -- & -- & -- & -- & 758 & 12m21s\\
8 & Contact\_SDA\_EncPIN\_Offline & -- & -- & -- & -- & 761 & 11m36s\\
\hline
9 & Contact\_DDA\_PlainPIN\_Online & \verified{} & \falsified{}$^{(2)}$ & 
\falsified{}$^{(1,2)}$ & \falsified{}$^{(1)}$ & 766 & 13m48s\\
10 & Contact\_DDA\_PlainPIN\_Offline & \verified{} & \falsified{}$^{(2)}$ & 
\falsified{}$^{(1,2)}$ & \falsified{}$^{(1)}$ & 769 & 12m20s\\
11 & Contact\_DDA\_OnlinePIN\_Online & \verified{} & \falsified{}$^{(2)}$ & 
\falsified{}$^{(2)}$ & \verified{} & 775 & 16m04s\\
12 & Contact\_DDA\_OnlinePIN\_Offline & -- & -- & -- & -- & 739 & 12m27s\\
13 & Contact\_DDA\_NoPIN\_Online & \verified{} & \falsified{}$^{(2)}$ & 
\falsified{}$^{(2)}$ & \verified{} & 769 & 12m15s\\
14 & Contact\_DDA\_NoPIN\_Offline & \verified{} & \falsified{}$^{(2)}$ & 
\falsified{}$^{(2)}$ & \verified{} & 772 & 8m43s\\
15 & Contact\_DDA\_EncPIN\_Online & \verified{} & \falsified{}$^{(2)}$ & 
\falsified{}$^{(1,2)}$ & \falsified{}$^{(1)}$ & 766 & 14m07s\\
16 & Contact\_DDA\_EncPIN\_Offline & \verified{} & \falsified{}$^{(2)}$ & 
\falsified{}$^{(1,2)}$ & \falsified{}$^{(1)}$ & 769 & 12m59s\\
\hline
17 & Contact\_CDA\_PlainPIN\_Online & \verified{} & \verified{} & 
\falsified{}$^{(1)}$ & \falsified{}$^{(1)}$ & 763 & 1h55m31s\\
18 & Contact\_CDA\_PlainPIN\_Offline & \verified{} & \verified{} & 
\falsified{}$^{(1)}$ & \falsified{}$^{(1)}$ & 766 & 14m10s\\
19 & \bfseries Contact\_CDA\_OnlinePIN\_Online & \verified{} & \verified{} & 
\verified{} & \verified{} & 781 & 6h03m05s\\
20 & Contact\_CDA\_OnlinePIN\_Offline & -- & -- & -- & -- & 739 & 12m15s\\
21 & \bfseries Contact\_CDA\_NoPIN\_Online & \verified{} & \verified{} & 
\verified{} & \verified{} & 775 & 2h31m23s\\
22 & \bfseries Contact\_CDA\_NoPIN\_Offline & \verified{} & \verified{} & 
\verified{} & \verified{} & 778 & 12m16s\\
23 & Contact\_CDA\_EncPIN\_Online & \verified{} & \verified{} & 
\falsified{}$^{(1)}$ & \falsified{}$^{(1)}$ & 763 & 1h59m44s\\
24 & Contact\_CDA\_EncPIN\_Offline & \verified{} & \verified{} & 
\falsified{}$^{(1)}$ & \falsified{}$^{(1)}$ & 766 & 14m00s\\
\hline\hline
\multicolumn{8}{l}{\bfseries Legend:}\\
\multicolumn{8}{l}{\verified{}: property verified\ \ \ \falsified{}: property 
falsified\ \ \  --: not applicable}\\
\multicolumn{8}{l}{(1): disagrees with the card on the \acrshort{CVM} used\ \ \ 
(2): disagrees with the card on the last \acrshort{AC}}
\end{tabular}
\end{table*}



Our analysis revealed disagreement, both between the 
terminal and the card and between the bank and the card, on the selected 
\acrshort{CVM} for transactions using \acrshort{SDA} or offline (plain or 
enciphered) PIN verification (Table~\ref{tab:results-contact}, Remark 1). %

For transactions where the terminal performed offline PIN verification, our 
analysis identifies a trace that represents the PIN bypass attack first 
observed by Murdoch \etal{}~\cite{MurdochDAB10} for transactions using 
\acrshort{SDA}. In this attack, a man-in-the-middle sends the
\emph{success} response to the terminal's PIN verification request. The 
actual request is blocked and so the card believes that no PIN 
verification was required, ergo the disagreement between the terminal and the 
card. The terminal forwards the transaction to the bank (either for online 
authorization or to collect the funds), which then leads to the disagreement 
between the bank and the card.

A prerequisite for this attack to succeed is that, even if the 
terminal sends to the card the \acrfull{CVMR} data object, which encodes the 
terminal's view of the \acrshort{CVM} used, 
and the card detects the mismatch with its own view of the \acrshort{CVM} used, 
the card does \emph{not} abort the transaction. This appears to be the case in 
practice (although EMV's specification is not explicit about this) and has been 
successfully tested with three different Mastercard cards using our Android 
app. Such tests, even though they were conducted contactless, give us a fair 
degree of confidence that it also occurs with contact transactions.

Our analysis also exhibits that all transactions using \acrshort{SDA} or 
\acrshort{DDA} are vulnerable to a \acrfull{TC} modification. This is because 
in neither of these methods the card authenticates the \acrshort{TC} 
to the terminal (Table~\ref{tab:results-contact}, Remark 2).%

%
%
In terms of secrecy, the results are identical for all models. The keys 
(private and shared) are secret, whereas the \acrshort{PAN} is not. 
Interestingly, our analysis reports that the PIN is not secret. 
%
A %
man-in-the-middle attack between the card and the terminal can use a 
compromised bank's private key to produce the card records needed to make the 
terminal believe that the only \acrshort{CVM} the card supports is plain PIN. 
These (fake) records are twofold: a list of supported \acrshort{CVM}s composed 
of plain 
PIN only, and either an \acrshort{SSAD} or a card's PKI certificate validating 
such a \acrshort{CVM} list. The terminal thus downgrades to plain PIN 
verification and consequently the PIN entered by the cardholder can be 
intercepted and learned by the attacker. This is a non-trivial attack though, 
as carrying this out in practice requires that the attacker:
\begin{enumerate}
\item knows a compromised bank's private key, and
\item inconspicuously controls the terminal's contact interface.
\end{enumerate}
Our model considers these two conditions to be possible, at least in theory. 
However, in practice they are difficult to achieve. We note however that a 
single compromised bank is sufficient, and it need not be the one that issued 
the victim's card. 

\subsubsection*{Summary}

We show that only three configurations of the EMV contact protocol guarantee 
secure transactions in terms of the three main properties we considered. These 
configurations all use \acrshort{CDA} as the authentication 
method and are typeset in bold in Table~\ref{tab:results-contact}. In 
combination with online PIN as the \acrlong{CVM}, the 
resulting target configuration allows all transactions (high and low value) and 
is secure.
It is also the only one of the three configurations that effectively checks
that the person presenting the card is the legitimate cardholder. 
%
The other two configurations instead delegate this check to the cashier, e.g., 
by a paper signature (whose actual verification is out of EMV's scope).
This makes these two configurations not usable for high-value transactions in 
many countries.
%

%

\subsection{Analysis Results for the EMV Contactless Protocol}
\label{sec:results-contactless}

%

Our analysis results for the 16 configurations of the EMV contactless protocol 
are summarized in Table~\ref{tab:results}. Here Tamarin uncovered new, 
potentially high-valued attacks.

\begin{table*}
\caption{Analysis Results for the EMV contactless protocol. All target models 
have 60 rules. Here we used the same computing setup as with the experiments 
shown in 
Table~\ref{tab:results-contact}. Again, the model(s) for which all four 
properties were verified are highlighted in bold.}
\label{tab:results}
\renewcommand{\arraystretch}{1.35}
\centering
\begin{tabular}{l l c c c c c r}
\hline\hline
\multirow{2}{*}{\bfseries No.} & 
\multirow{2}{*}{\bfseries Target model} & 
\multicolumn{4}{c}{\bfseries Properties}
& \multirow{2}{*}{\bfseries LoC}
& \multirow{2}{*}{\bfseries Time}
\\\cline{3-6}
& & \bfseries \itshape executable & \bfseries \itshape bank accepts & \bfseries 
\itshape auth. to terminal & \bfseries \itshape auth. to bank & &
\\\hline
1 & Visa\_EMV\_Low & \verified{} & \verified{} & \falsified{}$^{(1)}$ & 
\falsified{}$^{(1)}$ & 823 & 1m26s\\
2 & Visa\_EMV\_High & \verified{} & \verified{} & \falsified{}$^{(1)}$ & 
\falsified{}$^{(1)}$ & 823 & 1m30s\\
3 & Visa\_DDA\_Low & \verified{} & \falsified{}$^{(2)}$ & \falsified{}$^{(2)}$ 
& \verified{} & 832 & 31m41s\\
4 & \bfseries Visa\_DDA\_High & \verified{} & \verified{} & \verified{} & 
\verified{} & 841 & 25m00s\\
\hline
5 & Mastercard\_SDA\_OnlinePIN\_Low & \verified{} & \falsified{}$^{(2)}$ & 
\falsified{}$^{(2)}$ & \verified{} & 831 & 4m22s\\
6 & \bfseries Mastercard\_SDA\_OnlinePIN\_High & \verified{} & \verified{} & 
\verified{} & \verified{} & 840 & 12m28s\\
7 & Mastercard\_SDA\_NoPIN\_Low & \verified{} & \falsified{}$^{(2)}$ & 
\falsified{}$^{(2)}$ & \verified{} & 825 & 4m11s\\
8 & Mastercard\_SDA\_NoPIN\_High & --$^{(3)}$ & -- & -- & -- & 793 & 43s\\
9 & Mastercard\_DDA\_OnlinePIN\_Low & \verified{} & \falsified{}$^{(2)}$ & 
\falsified{}$^{(2)}$ & \verified{} & 837 & 8m18s\\
10 & \bfseries Mastercard\_DDA\_OnlinePIN\_High & \verified{} & \verified{} & 
\verified{} & \verified{} & 846 & 27m08s\\
11 & Mastercard\_DDA\_NoPIN\_Low & \verified{} & \falsified{}$^{(2)}$ & 
\falsified{}$^{(2)}$ & \verified{} & 831 & 8m11s\\
12 & Mastercard\_DDA\_NoPIN\_High & --$^{(3)}$ & -- & -- & -- & 799 & 47s\\
13 & \bfseries Mastercard\_CDA\_OnlinePIN\_Low & \verified{} & \verified{} & 
\verified{} & \verified{} & 846 & 19m44s\\
14 & \bfseries Mastercard\_CDA\_OnlinePIN\_High & \verified{} & \verified{} & 
\verified{} & \verified{} & 846 & 47m21s\\
15 & \bfseries Mastercard\_CDA\_NoPIN\_Low & \verified{} & \verified{} & 
\verified{} & \verified{} & 840 & 18m38s\\
16 & Mastercard\_CDA\_NoPIN\_High & --$^{(3)}$ & -- & -- & -- & 799 & 49s\\
\hline\hline
\multicolumn{8}{l}{\bfseries Legend:}\\
\multicolumn{8}{l}{\verified{}: property verified\ \ \ \falsified{}: property 
falsified\ \ \ --: not applicable\ \ \ (1): disagrees with the card on the 
\acrshort{CVM} used}\\
\multicolumn{8}{l}{(2): disagrees with the card on the \acrshort{AC}\ \ \ (3): 
high-value transactions without \acrshort{CVM} are not completed over the 
contactless interface}
\end{tabular}
\end{table*}

Our analysis shows that the Mastercard contactless protocol provides security 
for all high-value transactions. 
During transactions using \acrshort{SDA} or \acrshort{DDA}, the card does not 
authenticate the \acrfull{AC} to the terminal (Table~\ref{tab:results}, Lines 
5, 7, 9, and 11, Remark 2). Therefore, during \emph{offline} transactions using 
either of these methods, a man-in-the-middle can modify the \acrshort{AC} (or 
\acrlong{TC} due to being offline), which the terminal accepts given that it 
cannot verify its correctness. The mismatching \acrshort{AC} will later be 
detected by the issuing bank. This violates both properties formalized in 
Definitions~\ref{def:bank-accepts} and~\ref{def:auth-to-terminal}.

\begin{figure*}
\centering
\includegraphics[scale=.53]{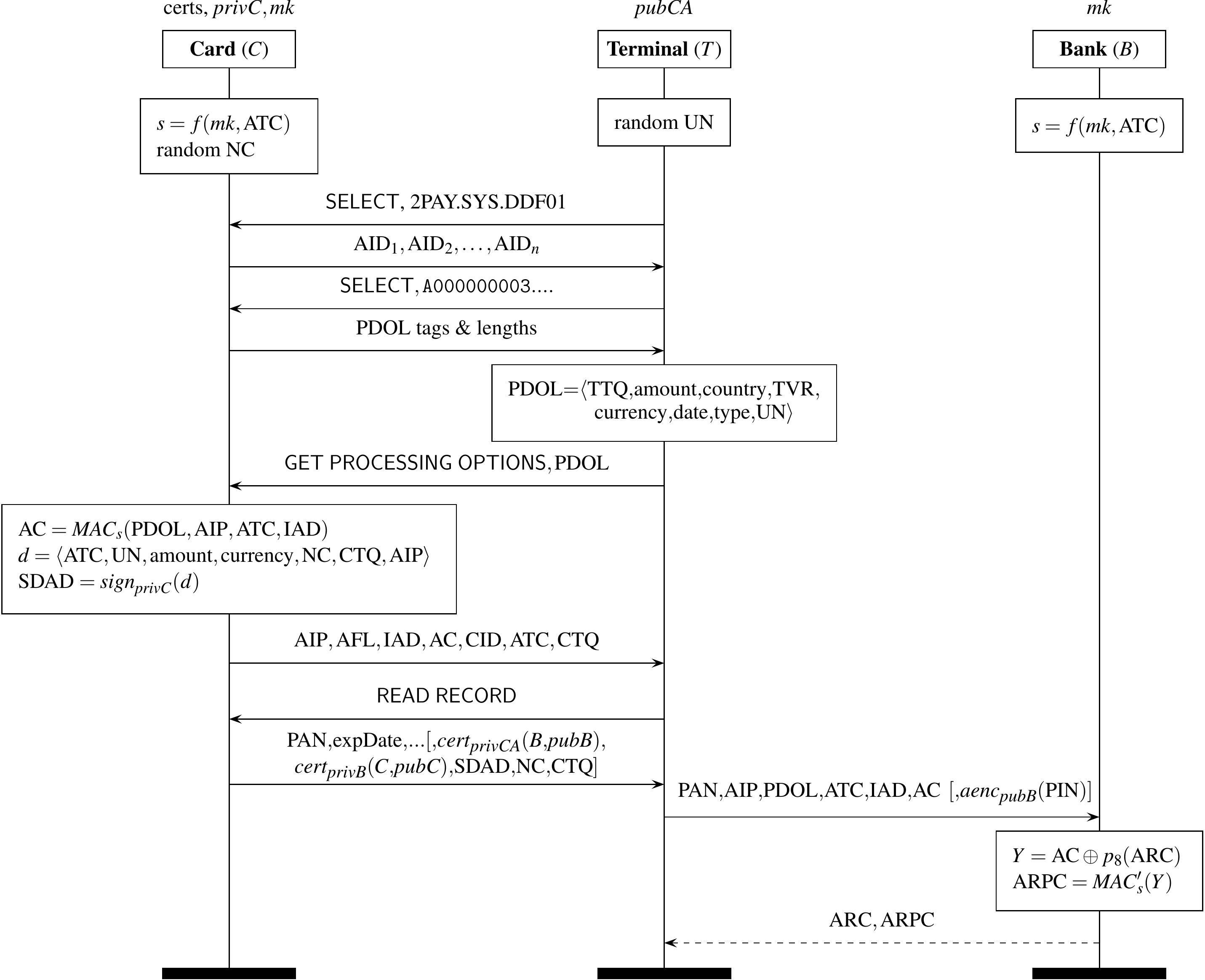}
\caption{The Visa contactless protocol. The terminal's request for cardholder 
verification and online authorization is encoded in the \acrshort{PDOL}, 
specifically in the \acrlong{TTQ} (\acrshort{TTQ}, tag \hexfont{9F66}). 
The card's response to the \acrshort{TTQ} requests is encoded in the 
\acrlong{CTQ} (\acrshort{CTQ}, tag \hexfont{9F6C}). The input to the 
\acrshort{AC} represented here includes the full \acrshort{PDOL} as 
per~\cite{security_and_key11}; proprietary cryptograms might use 
fewer data objects~\cite{GallowayY19}.}
\label{fig:visa}
\end{figure*}

To our surprise, all-but-one of the Visa contactless protocol's configurations 
fail to provide security. The Visa protocol is shown in Figure~\ref{fig:visa} 
and the secure configuration Visa\_DDA\_High corresponds to the protocol 
executions where all bracketed terms are present, especially the 
\acrshort{SDAD}. The insecure configuration Visa\_DDA\_Low, which applies to 
offline transactions, can be abused similarly to the aforementioned issue with 
Mastercard. 

Particularly critical are the violations of authentication in the 
configuration Visa\_EMV\_High (Table~\ref{tab:results}, Line 2, Remark 1), 
which applies to high-value transactions and corresponds to the protocol 
executions where the bracketed terms in the card's last message are not 
present. Our Tamarin analysis identifies a trace for an accepted 
transaction where neither the terminal nor the bank agree 
with the card on the \acrfull{CTQ}. The \acrshort{CTQ} is a card-sourced data 
object that tells the terminal which \acrshort{CVM} is to be used 
(see~\cite{visa19}, p. 97). The trace shows that, whereas the card's view 
of the \acrshort{CTQ} is a request for online PIN verification, the terminal's 
view indicates that the \acrfull{CDCVM} was performed, which makes the terminal 
consider the cardholder successfully verified (see~\cite{visa19}, pp. 68--69). 
This is possible because no cryptographic protection of the \acrshort{CTQ} is 
offered. This flaw is critical since it allows an attacker to bypass the PIN 
for a victim's Visa card, as pointed out in the introduction.

Bypassing the PIN by enforcing the \acrlong{CDCVM} in Mastercard transactions is 
not possible because the card's support for this \acrshort{CVM} is determined 
by the second bit of \acrshort{AIP}'s first byte. The \acrshort{AIP} 
is authenticated through the \acrshort{AC} and, if present, the \acrshort{SDAD} 
too. Consequently, modifying this data object would result in a declined 
transaction.

In terms of secrecy, the results are 
identical for all models and are as expected. The keys (private and shared) and 
the PIN are secret, whereas the \acrshort{PAN} is not. 

\subsubsection*{Summary}

Our analysis proves that Mastercard transactions using \acrshort{CDA} are 
secure. Fortunately, this is the most common kind of Mastercard transaction 
that is currently taking place. In contrast, critical flaws were found 
in common, currently used configurations of the Visa protocol. These flaws can 
be turned into practical attacks, which we describe in the next section.

\section{Attack and Defense}\label{sec:attack-and-fixes}

Our analysis of EMV's security uncovered numerous serious shortcomings. 
Particularly critical are the issues encountered in EMV contactless, because of 
their practical relevance given that tampering with the card-terminal 
contactless channel over \acrshort{NFC} is much simpler than tampering with 
this channel over the contact chip. In this section we show how these issues 
can be exploited by an attacker to carry out fraudulent transactions. We also 
suggest fixes that lead to verified, secure contactless transactions.

\subsection{Setup}

\begin{figure}
\centering
\begin{tikzpicture}
\node (reader) at (-3.9,0) {\includegraphics[scale=.05]{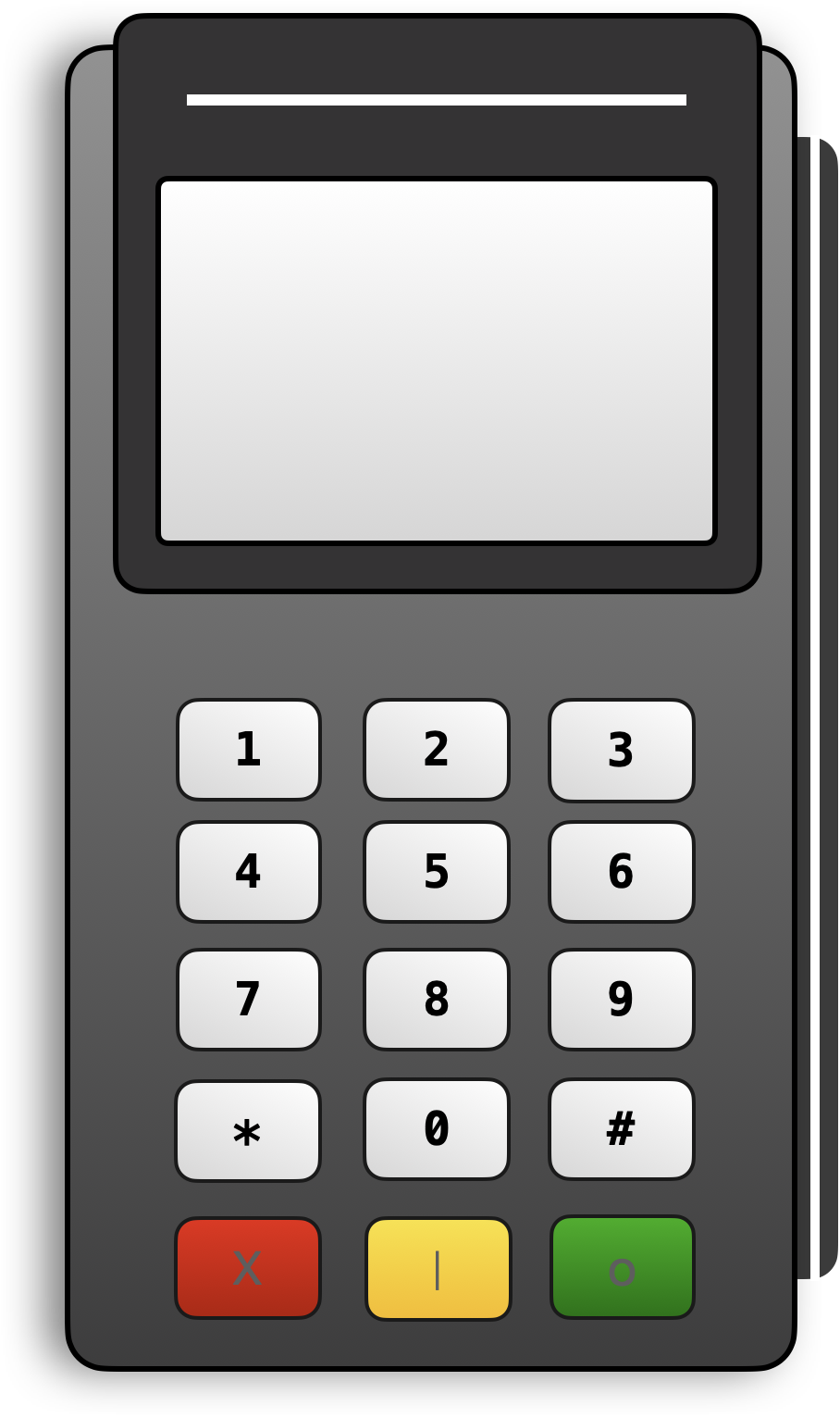}};
\node (proxytag) at (-2.5, 0) {\includegraphics[scale=.12]{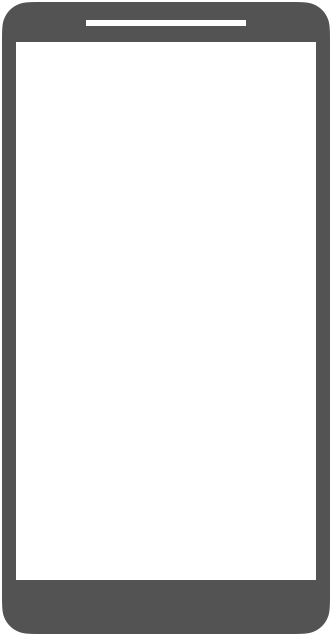}};
\node (proxyreader) at (2.5, 0) {\includegraphics[scale=.12]{phone.png}};
\node (tag) at (3.9,0) {\includegraphics[scale=.105]{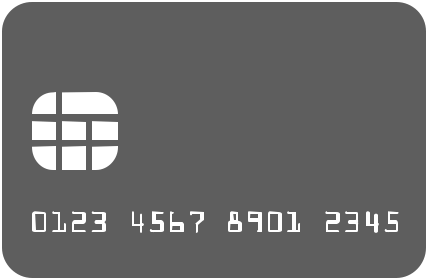}};

\node at (-3.9, .8) [above] {\circled{1}};
\node at (-2.5, .8) [above] {\circled{2}};
\node at (2.5, .8) [above] {\circled{3}};
\node at (3.9, .8) [above] {\circled{4}};

\node (WiFi) [rectangle, minimum width=4cm, minimum 
height=1.8cm, draw=gray, dotted, thick] at (0,.2) {};
\node at (0,1.1) [above] {WiFi};

\node at (0,0.38) [above] {\small \acrshort{APDU} commands};
\draw [->] (-1.8,0.4) -- (1.8,0.4);
\draw [<-] (-1.8,-0.4) -- (1.8,-0.4);
\node at (0,-0.52) [above] {\small \acrshort{APDU} responses};

\node at (-3.18,-0.3) [below] {\tiny \acrshort{NFC}};
\draw [bend left] (-3.3, 0.1) edge (-3.3, -0.1);
\draw [bend left] (-3.2, 0.2) edge (-3.2, -0.2);
\draw [bend left] (-3.1, 0.3) edge (-3.1, -0.3);

\node at (3.18,-0.3) [below] {\tiny \acrshort{NFC}};
\draw [bend left] (3.05, 0.1) edge (3.05, -0.1);
\draw [bend left] (3.15, 0.2) edge (3.15, -0.2);
\draw [bend left] (3.25, 0.3) edge (3.25, -0.3);
\end{tikzpicture}
\caption{A relay attack on contactless payment, where (1) 
is a payment terminal, (4) is a contactless card, and the attacker's equipment 
are the devices (2) and (3), which are the card emulator and the \acrshort{POS} 
emulator, respectively.}
\label{fig:relay}
\end{figure}

We developed a proof-of-concept Android application to demonstrate 
the practical impact of the shortcomings uncovered by our formal analysis. Our 
application supports man-in-the-middle attacks on 
top of a \emph{relay attack}~\cite{FrancisHMM11,SportielloC13,ChothiaGRBT15} 
architecture, depicted in Figure~\ref{fig:relay}. In this architecture, the 
attacker employs two mobile devices: one running our app in \emph{\acrfull{POS} 
emulator} mode and the other in \emph{card emulator} mode. Both devices must 
have \acrshort{NFC} support and run Android 4.4 KitKat (API level 19) or later. 
The card emulator device must support Android's \acrfull{HCE}~\cite{hce}.

To conduct the attacks, the \acrshort{POS} emulator must be held near the card 
to be attacked and the card emulator must be held near the payment terminal. 
The two emulators communicate wirelessly through a TCP/IP socket channel over 
WiFi. A man-in-the-middle attack modifies, as appropriate:
\begin{itemize}
\item the inbound commands read from the wireless channel before delivering 
them to the card through the \acrshort{NFC} channel, and
\item the card's responses before transmitting them to the card emulator 
through the WiFi channel.
\end{itemize}

\subsection{Bypassing Cardholder Verification}\label{sec:pin-attack}

In a Visa contactless transaction, the card's response to the terminal's 
\cmdfont{GET~PROCESSING~OPTIONS} command carries the \acrfull{CTQ}. The 
\acrshort{CTQ} is a 2-byte data object that 
instructs the terminal which \acrfull{CVM} is to be used. As explained in 
Section~\ref{sec:results-contactless}, our analysis revealed that the 
card authenticates the \acrshort{CTQ} neither to the terminal nor to the bank 
(Table~\ref{tab:results}, Line 2, Remark 1). Our app 
exploits this and implements a man-in-the-middle attack that:
\begin{itemize}
\item \textbf{clears the 8th bit of \acrshort{CTQ}'s first byte}, which tells 
the terminal that online PIN verification is not required; and
\item \textbf{sets the 8th bit of \acrshort{CTQ}'s second byte}, which tells 
the terminal that the \acrlong{CDCVM} was performed.
\end{itemize}

\begin{figure}
\centering
\begin{tabular}{c c}
\includegraphics[scale=.078]{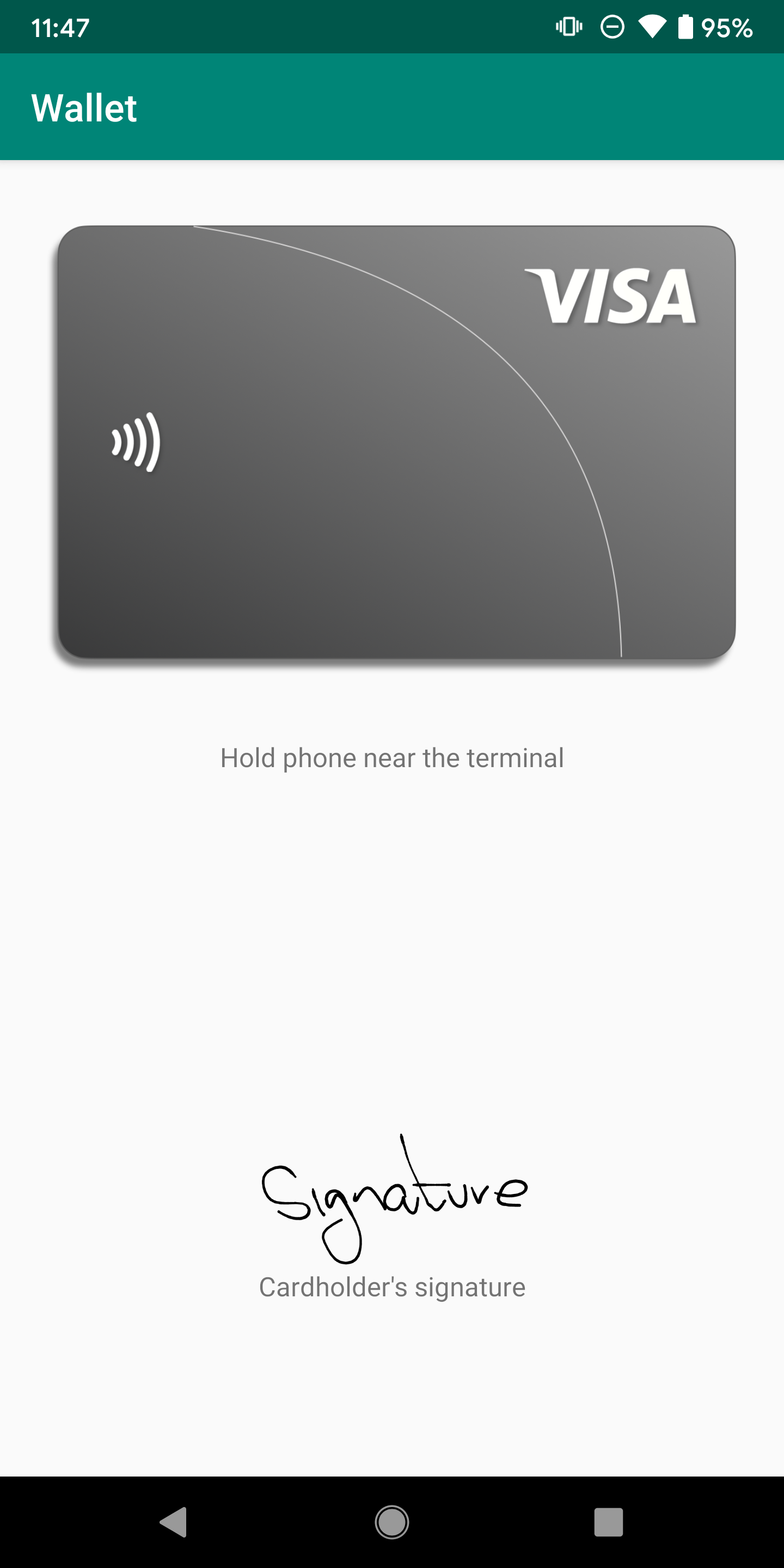}&
\includegraphics[scale=.078]{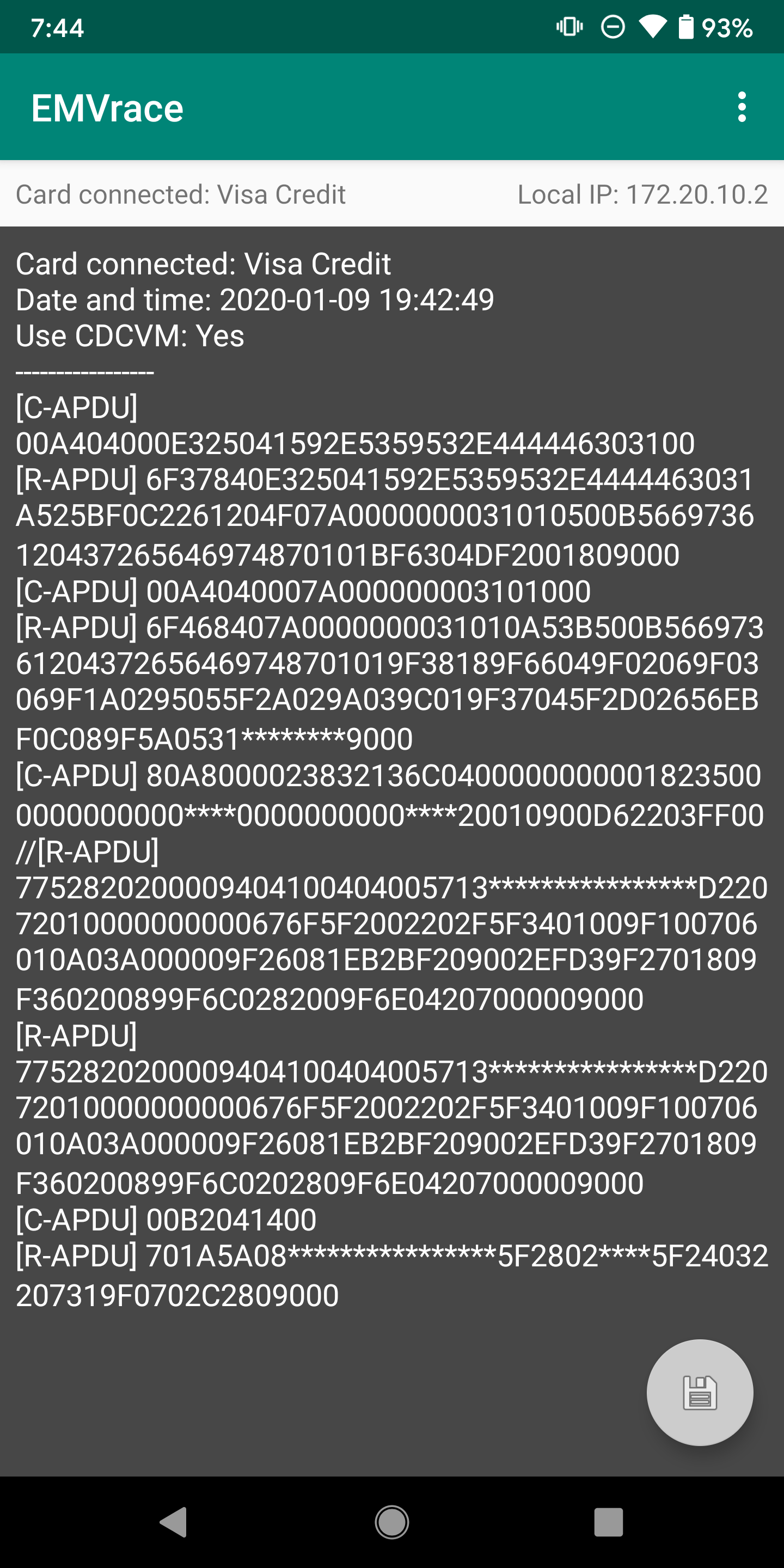}\\
(a) Card emulator & (b) POS emulator
\end{tabular}
\caption{Screenshots of our app. The card emulator may display the cardholder's 
signature, in which case it should match the attacker's signature. The log 
displayed in the POS emulator corresponds to a real transaction of 182.35 Swiss 
Francs (CHF).}
\label{fig:screenshhots}
\end{figure}

Using our app, we have successfully carried out a number of real-world, 
PIN-less transactions with amounts greater than the domestic 
\acrshort{CVM}-required limit with Visa credit and debit cards. 
Figure~\ref{fig:screenshhots} shows screenshots of our app. The transaction log 
displayed in the POS emulator screen in this figure corresponds to one of such 
transactions. A video demonstration of the attack for a 200 CHF transaction is 
available at~\cite{emvrace}.

Our attack should also work for the 
EMV Contactless Kernels 6~\cite{discover19} (Discover) and 
7~\cite{unionpay19} 
(UnionPay), but these have not been tested yet. %
To avoid defrauding others, all of our tests were carried out with 
our own debit/credit cards, and in all attacks the purchased goods were paid 
for in full.

As discussed in Section~\ref{sec:related-work}, Galloway and 
Yunusov~\cite{GallowayY19} recently presented at BlackHat Europe another 
man-in-the-middle attack that also bypasses Visa's PIN verification. In contrast
to our PIN bypass attack, their attack does not clear the 8th bit of 
\acrshort{CTQ}'s first byte. Instead, it clears the 7th bit of the \acrlong{TTQ}' 
second byte. This bit tells the card whether the terminal requires 
cardholder verification for the transaction (see~\cite{visa19}, p. 115).

The \acrfull{TTQ} is a terminal-sourced data object passed onto the card 
within the payload of the \cmdfont{GET~PROCESSING~OPTIONS} command. The 
\acrshort{TTQ} is part of the \acrfull{PDOL} and, according to the EMV 
Security and Key Management book~\cite{security_and_key11} (p. 88), the 
\acrfull{AC} is a \acrshort{MAC} computed on the data referenced by the card's 
data object lists such as the \acrshort{PDOL}. For this reason, our Tamarin 
analysis does not report the attack of~\cite{GallowayY19} since the (generic) 
\acrshort{AC} should prevent the modification of the \acrshort{PDOL} and of the 
\acrshort{TTQ} in particular. Visa's proprietary \acrshort{AC} does not, as noted 
in~\cite{GallowayY19}. Clearly, our attack works even if the \acrshort{TTQ} is 
authenticated as it needs no modification.

Another noticeable difference between our attack and that of~\cite{GallowayY19} 
is on the implementation side. Their attack prototype is 
composed of two wired Raspberry Pi boards. This setup is rather conspicuous and 
could not be easily used outside of a lab environment. In contrast, our 
proof-of-concept implementation is an innocent-looking phone app that can, and 
has been, easily used in live, attended terminals. Moreover, as opposed to 
Galloway and Yunusov's attack, ours does not require that the card and the 
payment terminal are physically close. In fact, one can extend our app so that 
the relay channel covers even overseas distances. Surprisingly, Visa has shown 
no intention to fix such vulnerabilities, as noted in~\cite{GallowayY19}.

Observe that our attack, as well as that of~\cite{GallowayY19}, presume that 
the attacker's device is physically within \acrshort{NFC} proximity of the 
victim's card. 
%
These attacks can therefore be carried out 
by acquiring the actual card (e.g., stealing it or finding it if lost) or by 
holding the POS emulator near the card in the victim's possession.

\subsection{Unauthenticated Offline Transactions}
\label{sec:offline-attack}

For all low-value transactions of Visa as well as Mastercard with either 
\acrshort{SDA} or \acrshort{DDA} offline authentication, our Tamarin analysis 
uncovers a trace that violates the property that the bank accepts all 
terminal-accepted transactions (Table~\ref{tab:results}, Remark 2). The trace 
represents a transaction where the attacker modifies the \acrfull{TC} before 
delivering it to the terminal. The terminal reaches a state where the 
transaction is accepted given that the \acrfull{SDAD}, if produced and returned 
by the card, passes the terminal's verification. However, the issuing bank 
should later decline the transaction due to the wrong \acrshort{TC}. Recall 
that the terminal can only verify the correctness of the \acrshort{SDAD} but 
not of the \acrshort{TC} since the latter is verified using a symmetric key 
only known to the card and the bank.

This constitutes a ``free lunch'' attack in that the criminal can purchase 
low-value goods or services without actually being charged at all. This 
however is unlikely to be an attractive business model for criminals for 
two reasons. First, the fraudulent transactions are of low value. Second, 
the criminal's bank will likely not ignore the defrauded merchant's complaints 
indefinitely. 
For ethical reasons, we did not test this attack as it would constitute actual 
fraud.

\subsection{Defenses against Attacks on Visa}\label{sec:fixes}

As reported in Section~\ref{sec:results-contactless}, the most common 
configuration of the Mastercard contactless protocol in current use (namely 
\acrshort{CDA} in conjunction with online PIN) is secure. Visa's 
configurations, on the other hand, are not. Fortunately, Visa's problems can be 
fixed by implementing the three changes that we describe next. These changes 
can be realized by Visa and the banks in a reasonable amount of time and 
effort, without affecting those cards currently in circulation.

The Visa contactless protocol~\cite{visa19} specifies that special-purpose 
readers may perform \acrfull{DDA} for online transactions. This is 
indeed the only configuration of this protocol where all three 
security properties hold (Table~\ref{tab:results}, Line 4). This is 
not a common configuration though, as indicated by our tests. We performed 
tests on over ten different live terminals at different merchants, and none of 
them used this configuration. Therefore, to prevent the PIN bypass attack 
described in Section~\ref{sec:pin-attack}, we recommend that terminals should 
use \acrshort{DDA} for online transactions. That is, all the terminals 
must, for all transactions:
\begin{enumerate}
\item set the first bit of \acrshort{TTQ}'s first byte, and
\item verify the \acrshort{SDAD}.
\end{enumerate}


If implemented, these two measures would require high-value transactions to be 
processed with Visa's secure configuration. This is of course assuming that the 
cards used for such transactions are capable of producing digital signatures, 
which modern cards are. Furthermore, to prevent the offline attack of 
Section~\ref{sec:offline-attack}, we propose that either:
\begin{enumerate}[label={3\alph{enumi})}]
\item all terminals set the 8th bit of \acrshort{TTQ}'s second byte for all 
transactions; or
\item $\left\langle \nc, \CID, \AC, \PDOL, \ATC, \CTQ, \UN, \IAD, 
\AIP\right\rangle$ is the input to the \acrshort{SDAD}, i.e., $d$ in 
Figure~\ref{fig:visa}.
\end{enumerate}

The fix 3(a) requires all transactions to be processed online and is preferable 
over 3(b) because 3(a) does not require changes to the standard and it 
therefore does not affect the consumer cards in circulation. Furthermore, 
offline transactions are not presently common; none of the more than 30 
transactions we 
carried out during our tests were authorized offline. However, if the 
capability to process certain transactions offline is imperative (e.g., in 
transit systems or street parking meters) then more aggressive fixes would be 
needed such as that of 3(b).

We have verified the three fixes recommended here. Together, they defend 
against the attacks reported in this paper as well as any other 
attacks that derive from violations of the considered security 
properties. These fixes, except for 3(b), can be deployed on the terminals' 
software and so they are attractive in terms of implementation because 
software updates on terminals should be significantly less expensive 
and faster than other, more aggressive actions such as blocking cards in 
circulation and issuing new ones. 

\section{Conclusions}\label{sec:conclusion}

We have presented a formal model of the latest version of the EMV standard that 
features all relevant methods for offline data authentication, cardholder 
verification, and transaction authorization. Using the Tamarin tool, we 
conducted a full-scale, automatic, formal analysis of this model, uncovering 
numerous security flaws. These flaws violate fundamental security properties 
such as authentication and other guarantees about accepted transactions.  We 
also used our model to identify EMV configurations that lead to secure 
transactions, and proved their correctness. 

Our analysis revealed surprising differences between the
security of the contactless payment protocols of Mastercard and Visa, 
showing that Mastercard is more secure than Visa. We found no major 
issues with the Mastercard protocol version running in 
modern cards. Our analysis revealed only minor shortcomings arising from older 
authentication modes (\acrshort{SDA} and \acrshort{DDA}) that seem hard to 
exploit in practice. In contrast, Visa suffers from several critical issues. 
The shortcomings we report on lead to serious, practical attacks, including a 
PIN bypass for transactions that surpass the cardholder verification limit. 
Using our proof-of-concept Android application, we successfully tested this 
attack on real-world transactions in actual stores. Our attack shows that the 
PIN is useless for Visa contactless transactions. %
As a result, in our view, the liability shift from banks to consumers or 
merchants is unjustified for such transactions: Banks, EMVCo, Visa, or some 
entity other than the consumer or merchant should be liable for such 
fraudulent transactions.

As part of our analysis we suggested and verified fixes that banks and Visa
can deploy on existing terminals to prevent current and future
attacks.  The good news is that these fixes do not require changes to the EMV 
standard itself or to consumer cards currently in circulation and they can 
therefore be feasibly deployed by software updates.

As future work, we plan to merge our EMV contact and contactless models into 
a single model in order to analyze cross-protocol executions. We also plan to 
further refine and extend our models to take into consideration other, 
possibly stronger, adversaries such as adversaries with dynamic compromise 
capabilities.

\bibliographystyle{ieeetr}
\bibliography{emv}

\printglossary[type=\acronymtype]

\appendices

\section{Target Model Generation}
\label{app:target-models-generation}

We construct the \emph{target models} from the rules of a generic model as well 
as additional rules that produce the \fact{Commit} facts used 
for the (in)validation of properties. We have written a \verb|Makefile| 
script 
that generates the target models by instantiating the following variables:
\begin{itemize}
\item \verb|generic| defines the generic model. Valid instances are:
\begin{itemize}
\item \verb|Contact|, and
\item \verb|Contactless|.
\end{itemize}
\item \verb|kernel| defines the kernel of the contactless transaction. Valid 
instances are:
\begin{itemize}
\item \verb|Mastercard|, and
\item \verb|Visa|.
\end{itemize}
\item \verb|auth| defines the Offline Data Authentication (ODA) method. Valid 
instances are:
\begin{itemize}
\item \verb|SDA|,
\item \verb|DDA|,
\item \verb|CDA|, and
\item \verb|EMV| (for contactless transactions only).
\end{itemize}
\item \verb|CVM| defines the cardholder verification method used/supported. 
Valid instances are:
\begin{itemize}
\item \verb|NoPIN|,
\item \verb|PlainPIN| (for contact transactions only),
\item \verb|EncPIN| (enciphered PIN, for contact transactions only), and
\item \verb|OnlinePIN|.
\end{itemize}
\item \verb|value| defines the value of the contactless transaction. Valid 
instances are:
\begin{itemize}
\item \verb|Low| (below the CVM-required limit), and
\item \verb|High| (above the CVM-required limit).
\end{itemize}
\item \verb|authz| defines the type of authorization of the contact 
transaction. Valid instances are:
\begin{itemize}
\item \verb|Offline|, and
\item \verb|Online|.
\end{itemize}
\end{itemize}

The execution of \verb|make| with a choice of variable instances determining a 
target configuration generates the target model and analyzes it with Tamarin. 
To understand how we instrument the actual target models 
auto-generation, consider the code snippet depicted in 
Figure~\ref{fig:snippet}, taken from our generic model of the EMV contactless 
protocol.

\begin{figure}
\begin{lstlisting}[language=Tamarin]
/*if(Visa)
rule Terminal_Commits_ARQC_Visa:
    let PDOL = <TTQ, $amount, country, currency,              date, type, ~UN>
        /*if(DDA) AIP = <'DDA', data> endif(DDA)*/
        /*if(EMV) AIP = <'EMV', data> endif(EMV)*/
        /*if(Low) value = 'Low' endif(Low)*/
        /*if(High) value = 'High' endif(High)*/
        transaction = <~PAN, AIP, CVM, PDOL, ATC,                    AC, IAD>
    in
    [ Terminal_Received_AC_Visa($Terminal, $Bank,
        $CA, nc, 'ARQC', transaction, ~channelID),
      !Value($amount, value),
      Recv($Bank, $Terminal,
        <~channelID, 'Visa', '2'>, <'ARC', ARPC>) ]
  --[ TerminalAccepts(transaction),
      Commit('Terminal', ~PAN,
        <'Card', 'Terminal', transaction>),
      Commit($Terminal, $Bank,
        <'Bank', 'Terminal', transaction>),
      Honest($CA), Honest($Bank),
      Honest($Terminal), Honest(~PAN)]->
    [ ]
endif(Visa)*/
\end{lstlisting}
\caption{Snippet from the EMV contactless protocol model.}
\label{fig:snippet}
\end{figure}

This piece of code is \emph{activated} (uncommented), and so the rule 
becomes part of the target model, if the target configuration includes 
\verb|kernel=Visa|. Furthermore, depending on the rest of the target 
configuration, the AIP and value are activated. For example, if our target 
configuration includes \verb|auth=DDA| and \verb|value=High|, then the rule 
becomes the one depicted in Figure~\ref{fig:snippet-uncommented}. This (new) 
rule models the terminal's acceptance of an online-authorized transaction and 
produces the corresponding \fact{Commit} and \fact{TerminalAccepts} facts.

\begin{figure}
\begin{lstlisting}[language=Tamarin]
rule Terminal_Commits_ARQC_Visa:
    let PDOL = <TTQ, $amount, country, currency,              date, type, ~UN>
        AIP = <'DDA', data>
        value = 'High'
        transaction = <~PAN, AIP, CVM, PDOL, ATC,                    AC, IAD>
    in
    [ Terminal_Received_AC_Visa($Terminal, $Bank,
        $CA, nc, 'ARQC', transaction, ~channelID),
      !Value($amount, value),
      Recv($Bank, $Terminal,
        <~channelID, 'Visa', '2'>, <'ARC', ARPC>) ]
  --[ TerminalAccepts(transaction),
      Commit('Terminal', ~PAN,
        <'Card', 'Terminal', transaction>),
      Commit($Terminal, $Bank,
        <'Bank', 'Terminal', transaction>),
      Honest($CA), Honest($Bank),
      Honest($Terminal), Honest(~PAN)]->
    [ ]
\end{lstlisting}
\caption{Snippet from the Visa\_DDA\_High target model.}
\label{fig:snippet-uncommented}
\end{figure}

\end{document}